\documentclass[aps,prd,twocolumn,superscriptaddress,showpacs,floatfix]{revtex4-2} 
\usepackage{dcolumn}
\usepackage{amsmath,amssymb,bm}
\usepackage{graphicx}
\usepackage{braket}
\usepackage{slashed,tensor}

\usepackage{newtxtext,newtxmath}
\usepackage{mathrsfs}
\usepackage{palatino}

\usepackage{color}
\usepackage{ifluatex}
\usepackage[utf8]{\ifluatex lua\fi inputenc}
\usepackage[T1]{fontenc}
\usepackage{epstopdf}

\usepackage{lineno}
\usepackage[FIGTOPCAP]{subfigure}
\usepackage{stackengine}

\usepackage{hyperref}
\hypersetup{colorlinks=true,citecolor=blue,linkcolor=blue,urlcolor=blue}

\usepackage{centernot}

\begin{document}
\title{Double-strangeness exchange reactions in a hybrid Regge-plus-resonance approach}
\author{Sang-Ho Kim}
\affiliation{Department of Physics and Origin of Matter and Evolution of Galaxy
(OMEG) Institute, Soongsil University, Seoul 06978, Korea}
\author{Jung Keun Ahn}
\email[E-mail: ]{ahnjk@korea.ac.kr}
\affiliation{Department of Physics, Korea University, Seoul 02841, Korea}
\author{Shin Hyung Kim}
\affiliation{Advanced Science Research Center, Japan Atomic Energy Agency, Tokai 319-1195, Japan}
\author{Seung-il Nam}
\affiliation{Department of Physics, Pukyong National University, Busan 48513, Korea}
\affiliation{Asia Pacific Center for Theoretical Physics, Pohang, Gyeongbuk 37673, Korea}
\author{Myung-Ki Cheoun}
\affiliation{Department of Physics and Origin of Matter and Evolution of Galaxy
(OMEG) Institute, Soongsil University, Seoul 06978, Korea}

\date{\today}

\begin{abstract}
We investigate double-strangeness exchange reactions, $K^-p\to K^+\Xi^-$ and $K^-p\to K^0\Xi^0$, 
using an effective Lagrangian approach based on a hybrid Regge-plus-resonance model 
involving rescattering diagrams. 
We consider the background processes that include $\Lambda$, $\Sigma$, and
$\Sigma(1385)$ Regge trajectories in the $u$ channel and the $s$-channel Born-term diagrams. 
The $s$-channel hyperon resonances account for the bump structures in the total cross sections. 
The $s$-channel $\Lambda(2100,7/2^-)$ and $\Sigma(2030,7/2^+)$ resonances contribute 
significantly, and the $\Sigma(2250)$ resonance favors $J^P = 7/2^-$.
The interference pattern between the $\Lambda(2100,7/2^-)$ and $\Sigma(2030,7/2^+)$
amplitudes is essential for describing forward differential cross sections
in the resonance region $2.07 < \sqrt{s} < 2.15$ GeV.
The absence of the $S=-2$ meson leads to the inclusion of the meson-baryon rescattering diagrams to
describe the $K^- p \to K\Xi$ reaction data, proving the significant contribution of 
the $(\phi,\rho,\omega)$--$(\Lambda,\Sigma)$ rescattering diagrams.
\end{abstract}
\maketitle

\section{Introduction} \label{Section:I}

Hadron spectroscopy provides essential information regarding Quantum Chromodynamics (QCD) 
in the nonperturbative regime \cite{Mai:2023}. In recent years, new hadrons have been observed 
owing to experimental advances, thus making the hadron spectrum abundant. 
Although the excited states of nucleons are relatively well known, 
only a limited amount of hyperon resonances are available for testing the lattice QCD calculation results. 
A complete set of hyperon resonances can drastically alter our understanding of the nonperturbative QCD 
and quark confinement \cite{Lutz:2016,Hyodo:2021}. 
Furthermore, owing to their large widths, it is challenging to disentangle 
$\Lambda$ and $\Sigma$ resonances overlapping near $\sqrt{s} = 2.0$ GeV. 
These high-mass hyperon resonances may couple selectively with the $K\Xi$ channels~\cite{Capstick:2000qj,Isgur:1978wd}.

The $\overline{K}N\to K\Xi$ reactions on the nuclear targets provide 
a unique test ground for the spectroscopy of double hypernuclear states and introduce a new means for studying
the elusive content of multiquark hadrons, such as the $H$-dibaryon. 
Double-strangeness exchange $(K^-, K^+)$ and $(K^-, K^0)$ reactions 
would benefit from the absence of a single-meson exchange in the $t$ channel. 
These reactions are crucial for imposing constraints on largely unknown vertex parameters 
for the decay of $\Lambda$ and $\Sigma$ resonances in the $\overline{K}N$ and $K\Xi$ channels.

The experimental data for the $(K^-, K^+)$ and $(K^-, K^0)$ reactions are primarily available 
from old bubble-chamber experiments in the sixties and seventies~
\cite{Carmony:1964zza,Birmingham:1966onr,London:1966zz,Berge:1966zz,Trippe:1967wat,
Burgun:1968ice,Trower:1968zz, Dauber:1969hg, SABRE:1971pzp,deBellefon:1972rjq,Carlson:1973td,
Rader:1973ja,Griselin:1975pa,Briefel:1977bp}, which placed
limitations on a rigorous and detailed theoretical treatment.
However, new insights into the double-strangeness exchange reactions 
have emerged from recent advances in experimental measurements. 
A long-term program for the $S=-2$ system study is ongoing at the J-PARC hadron experiment
facility \cite{Tamura:2012}.
A pilot experiment (E05) exploring $\Xi^-$-hypernuclear states reported a forward peaking 
in the differential cross section for the $K^- p \to K^+ \Xi^-$ reaction at 1.8 GeV$/c$
~\cite{Nagae:2019uzt,Gogami:2020}.
The high-statistics of the E05 data also demonstrated that the production cross section 
averaged over the forward angles reached a maximum value at 1.8 GeV$/c$. 
Recently, the $H$-dibaryon search experiment (E42) collected $(K^-, K^+)$ reaction data, 
providing $\Xi^-$ production cross-section and recoil polarization data in the forward region 
at 1.8 GeV$/c$ \cite{E42a,E42b}.
The $\Lambda$ and $\Sigma$ resonances are also critical for photoproduction 
and the $\overline{p}p\to\overline{\Xi}\Xi$ reactions. 
A recent experimental highlight is the measurement of $\Xi^-$ and $\Xi^\ast$ from photoproduction
at the Jefferson Lab with CLAS ~\cite{CLAS:2018kvn} and
GlueX detectors ~\cite{Ernst:2020vjd}. 
The GSI-FAIR facility intends to make advances 
in the $\overline{p}p\to\overline{\Xi}\Xi$ reactions with the PANDA detector~\cite{PANDA:2009yku}.
However, the $\gamma N$ and $\overline{p}p$ reactions require two steps to produce $\Xi/\overline{\Xi}$,
making it challenging to study $s$-channel hyperon resonances.

The forward $K^+/K^0$ angular distributions are not strong 
when double-charge or double-strangeness exchange is required in the $t$-channel. 
However, early experimental data demonstrated a sizable strength at forward angles 
for the double-strangeness exchange reactions. 
Therefore, several early theoretical efforts were made to elucidate 
forward peaking in the differential cross sections with single-particle exchange diagrams 
in the $s$- and $u$-channels and rescattering diagrams involving vector mesons and hyperons; 
however, these need to be more satisfactory when describing the data. 
Previous theoretical efforts include the effective Lagrangian
approach~\cite{Sharov:2011xq,Shyam:2011ys}, a unitarized chiral perturbation approach
~\cite{Feijoo:2015yja}, and coupled-channel approaches from global multi-channel analyses
by Argonne-Osaka~\cite{Kamano:2014zba}, Julish-Bonn-Washington~\cite{Landay:2018wgf}, and
Bonn-Gatchina~\cite{Matveev:2019igl}.
A phenomenological contact-term amplitude for the rescattering contribution
was included in the model-independent~\cite{Nakayama:2012zp,Jackson:2013bba} 
and model-dependent~\cite{Jackson:2015dva} analyses by using an effective Lagrangian approach.

This study investigates the double-strangeness exchange $(K^-, K^+)$ and $(K^-, K^0)$ reactions
in the tree-level effective Lagrangian approach based on a hybrid Regge-plus-resonance (RPR) model. 
Because the tree-level isobar models focus on selecting a relevant set of $s$-channel resonance
diagrams and Born terms involving non-resonant contributions~\cite{Sharov:2011xq,Shyam:2011ys,Jackson:2015dva}, 
there are several parameters to fit the data. 
In contrast, our model focuses on Reggeizing the Feynman diagrams 
involving the exchange of $S=-1$ hyperon trajectories in the $u$ channel~\cite{Storrow:1983ct}. 
The Regge formalism is well-suited for processes at high $s$ and small $|t|$ or $|u|$. 
We consider the $\Lambda$, $\Sigma$, and $\Sigma^\ast(1385)$ Regge trajectories in the
$u$-channel diagram.
The background of this model includes the $s$-channel Born-term 
and Reggeized $u$-channel contributions.
For the $s$-channel resonances, we consider the decay-branching fractions 
for the $\bar{K}N $ and $K\Xi$ channels from the Particle Data Group~(PDG)~\cite{PDG:2022pth}. 
The inclusion of the $s$-channel $\Lambda$ and $\Sigma$ resonances provides 
critical information regarding the vertex parameters of the $(K^-, K^+)$ and $(K^-, K^0)$ reactions. 
The absence of $S=-2$ mesons leads to a lack of the $t$-channel Born-term contribution. 
Instead, we consider rescattering diagrams mediated by pseudoscalar and vector mesons with 
$\Lambda$ and $\Sigma$ ground states, leading to an extension of the RPR model,
named the hybrid RPR model. 

In the next section, we outline the theoretical framework of the hybrid RPR model. Sec.~\ref{Section:II}A
focuses on the procedure for $u$-channel Reggeization of the high-energy amplitude. 
Sec.~\ref{Section:II}B presents our choice of $s$-channel hyperon resonances with deduced coupling
constants. Sec.~\ref{Section:II}C demonstrates the formalism of meson-baryon rescattering amplitudes.
The numerical results are presented in Sec.~\ref{Section:III}. Finally, the results are summarized in 
the last section.

\section{Theoretical Framework} \label{Section:II}

\begin{figure}[htb]
\includegraphics[width=6.5cm]{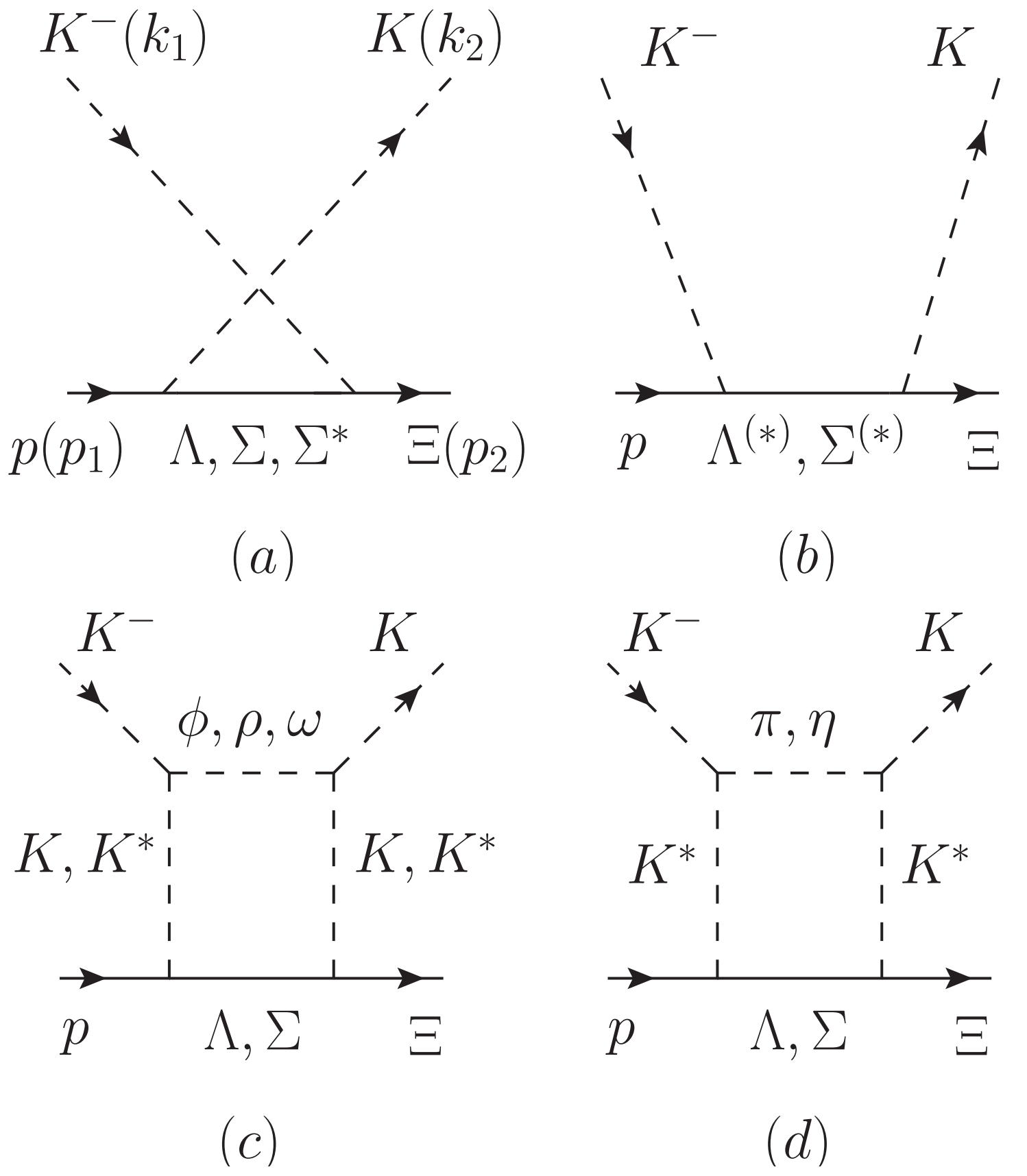}
\caption{Feynmann diagrams describing the $K^- p \to K \Xi$ reactions.
The intermediate states are (a) Reggeized $\Lambda$, $\Sigma$, and $\Sigma(1385)$ hyperons
in the $u$ channel; (b) ground and excited states of $\Lambda$ and $\Sigma$ hyperons in
the $s$ channel.
The rescattering diagrams of the $K^- p \to M_i B_i \to K \Xi$ processes are represented as shown in 
(c) $M_i=(\phi,\rho,\omega), B_i=(\Lambda,\Sigma)$ and
(d) $M_i=(\pi,\eta), B_i=(\Lambda,\Sigma)$.}
\label{FIG01}
\end{figure}

In this section, we introduce the theoretical framework for investigating the
$K^-+ p\to K+\Xi$
reactions using an effective Lagrangian approach.
We consider the $K^- p \to K^+ \Xi^-$ and $K^- p \to K^0 \Xi^0$ reactions.
The Feynman diagrams of the reactions are shown in Fig.~\ref{FIG01}, where
the four-momenta of the particles are denoted by $k_1$, $p_1$, $k_2$, and $p_2$.
Diagram (a) indicates an exchange of Reggeized $\Lambda$, $\Sigma$, and $\Sigma(1385)$
hyperons in the $u$ channel and diagram (b) denotes an $s$-channel process involving $\Lambda$ 
and $\Sigma$ ground states and their resonances with spin-parities up to $J^P = 7/2^\pm$ 
in the $s$-channel. 
The other diagrams represent the $K^- p \to M_i B_i \to K \Xi$ processes 
involving $M_i=(\phi,\rho,\omega)$ and $B_i=(\Lambda,\Sigma)$ by exchanging $K$ and $K^\ast$
in (c), and $M_i=(\pi,\eta)$ and 
$B_i=(\Lambda,\Sigma)$ with a $K^\ast$ exchange in (d).

We employ the following notation for the isodoublet fields:
\begin{align}
N = \left( \begin{array}{c} p \\ n \end{array} \right), \hspace{0.25em}
\Xi = \left( \begin{array}{c} \Xi^0 \\ \Xi^- \end{array} \right), \hspace{0.25em}
K = \left( \begin{array}{c} K^+ \\ K^0 \end{array} \right), \hspace{0.25em}
K' = \left( \begin{array}{c} \overline{K}^0 \\ -K^- \end{array} \right).
\end{align}
The effective Lagrangians for the exchanges of the spin-1/2 baryons (and their resonances)
can be expressed as follows:
\begin{align}
{\mathcal L}_{KN\Lambda}^{1/2^\pm} =& g_{KN\Lambda} \overline{\Lambda}
(D_{N\Lambda} \overline{K}) N + \rm{H.c.},                                                   \cr
{\mathcal L}_{K'\Xi\Lambda}^{1/2^\pm} =& g_{K\Xi\Lambda} \overline{\Xi}
(D_{\Lambda\,\Xi} K') \Lambda + \rm{H.c.},                                              \cr
{\mathcal L}_{KN\Sigma}^{1/2^\pm} =& g_{KN\Sigma} \overline{\mathbf\Sigma} \cdot
(D_{N\Sigma} \overline{K}) {\bf \tau} N + \rm{H.c.},                                         \cr
{\mathcal L}_{K'\Xi\Sigma}^{1/2^\pm} =& g_{K\Xi\Sigma } \overline{\Xi} {\bf \tau}
(D_{\Sigma\,\Xi} K') \cdot {\mathbf\Sigma} + \rm{H.c.},
\label{Lagrangians:1/2}
\end{align}
where
\begin{align}
D_{BB'} \equiv
\Gamma^{\pm} \left(- i\lambda + \frac{1-\lambda}{M_B + M_{B'}} \slashed \partial \right).
\label{AOperator}
\end{align}
Here, $\Gamma^+ \equiv \gamma_5$ and $\Gamma^- \equiv 1_{4 \times 4}$.
The pseudoscalar (PS) and pseudovector (PV) couplings correspond to $\lambda = 1$ and
$\lambda = 0$, respectively.

The effective Lagrangians corresponding to the exchanges of the
$J^P = (3/2^\pm,\,5/2^\pm,\, 7/2^\pm)$ $\Lambda$ resonances, which are respectively defined by
the following:
\begin{align}
{\mathcal L}_{KN\Lambda^\ast}^{3/2^\pm} =& \frac{g_{KN\Lambda^\ast}}{M_K} 
\overline{\Lambda}^{*\mu}
\Gamma^\mp (\partial_\mu \overline{K}) N + \rm{H.c.},                                     \cr
{\mathcal L}_{KN\Lambda^\ast}^{5/2^\pm} =& i \frac{g_{KN\Lambda^\ast}}{M_K^2} 
\overline{\Lambda}^{*\mu\nu}
\Gamma^\pm (\partial_\mu \partial_\nu \overline{K}) N + \rm{H.c.},                         \cr
{\mathcal L}_{KN\Lambda^\ast}^{7/2^\pm} =& - \frac{g_{KN\Lambda^\ast}}{M_K^3} 
\overline{\Lambda}^{*\mu\nu\alpha}
\Gamma^\mp (\partial_\mu \partial_\nu \partial_\alpha \overline{K}) N + \rm{H.c.},
\label{Lagrangians:KNL}
\end{align}
for the $K N \Lambda^\ast$ vertex;
\begin{align}
{\mathcal L}_{K'\Xi\Lambda^\ast}^{3/2^\pm} =& \frac{g_{K\Xi\Lambda^\ast}}{M_K} \overline{\Xi}\,
\Gamma^\mp (\partial_\mu K') \Lambda^{*\mu} + \rm{H.c.},                              \cr
{\mathcal L}_{K'\Xi\Lambda^\ast}^{5/2^\pm} =& i \frac{g_{K\Xi\Lambda^\ast}}{M_K^2} \overline{\Xi}\,
\Gamma^\pm (\partial_\mu \partial_\nu K') \Lambda^{*\mu\nu} + \rm{H.c.},               \cr
{\mathcal L}_{K'\Xi\Lambda^\ast}^{7/2^\pm} =& - \frac{g_{K\Xi\Lambda^\ast}}{M_K^3} \overline{\Xi}\,
\Gamma^\mp (\partial_\mu \partial_\nu \partial_\alpha K') \Lambda^{*\mu\nu\alpha} + \rm{H.c.},
\label{Lagrangians:KXiL}
\end{align}
for the $K' \Xi \Lambda^\ast$ vertex. The exchange of $\Sigma$ resonances is considered 
in the same manner as that in
Eq.~(\ref{Lagrangians:1/2}).

\subsection{Reggeized hyperon exchange in the $u$ channel}

\begin{figure*}[htb]
\stackinset{l}{-0.3cm}{t}{0.1cm}{(a)}{
\includegraphics[width=4.3cm]{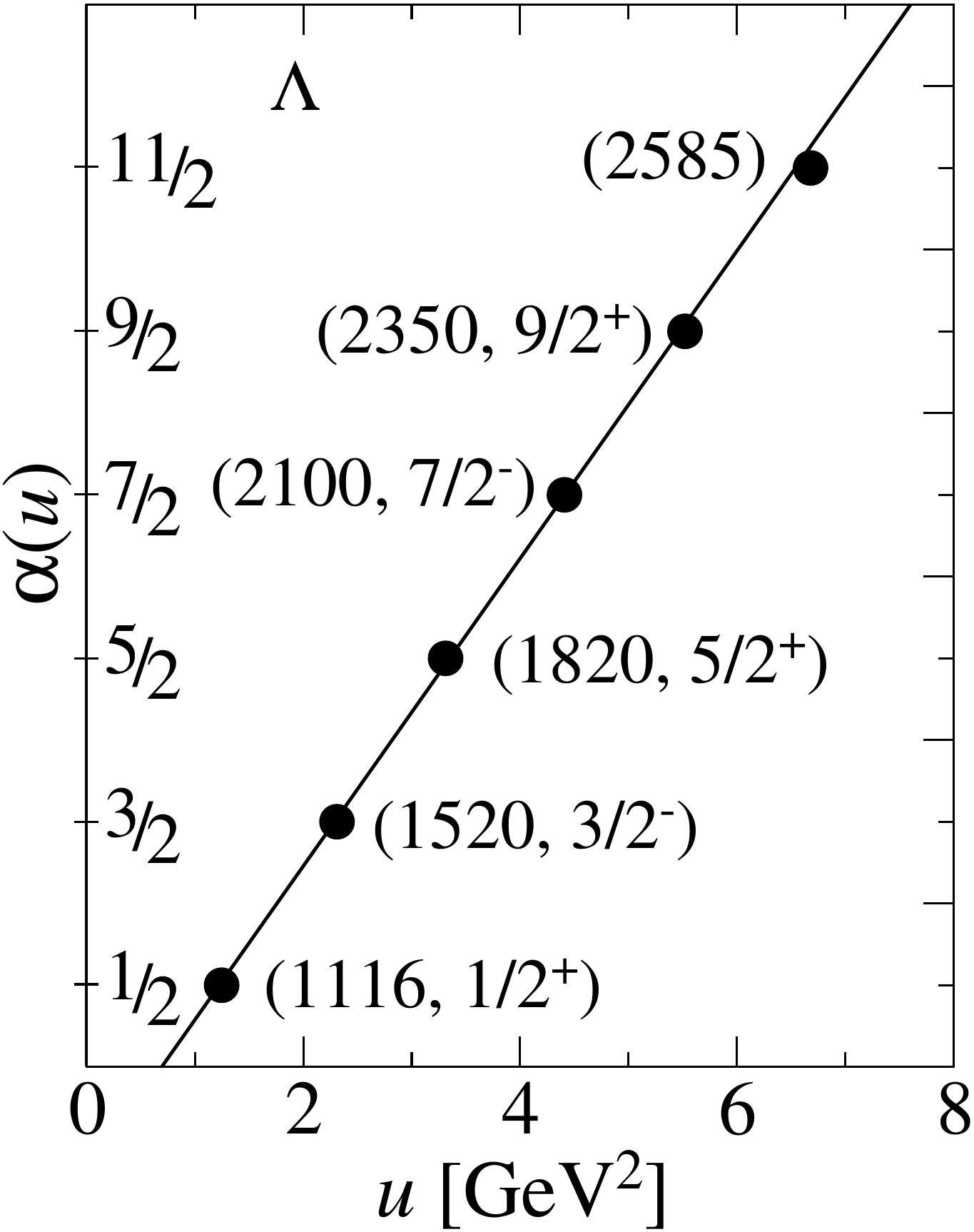}
}\hskip+2.5cm
\stackinset{l}{-0.3cm}{t}{0.1cm}{(b)}{
\includegraphics[width=7.5cm]{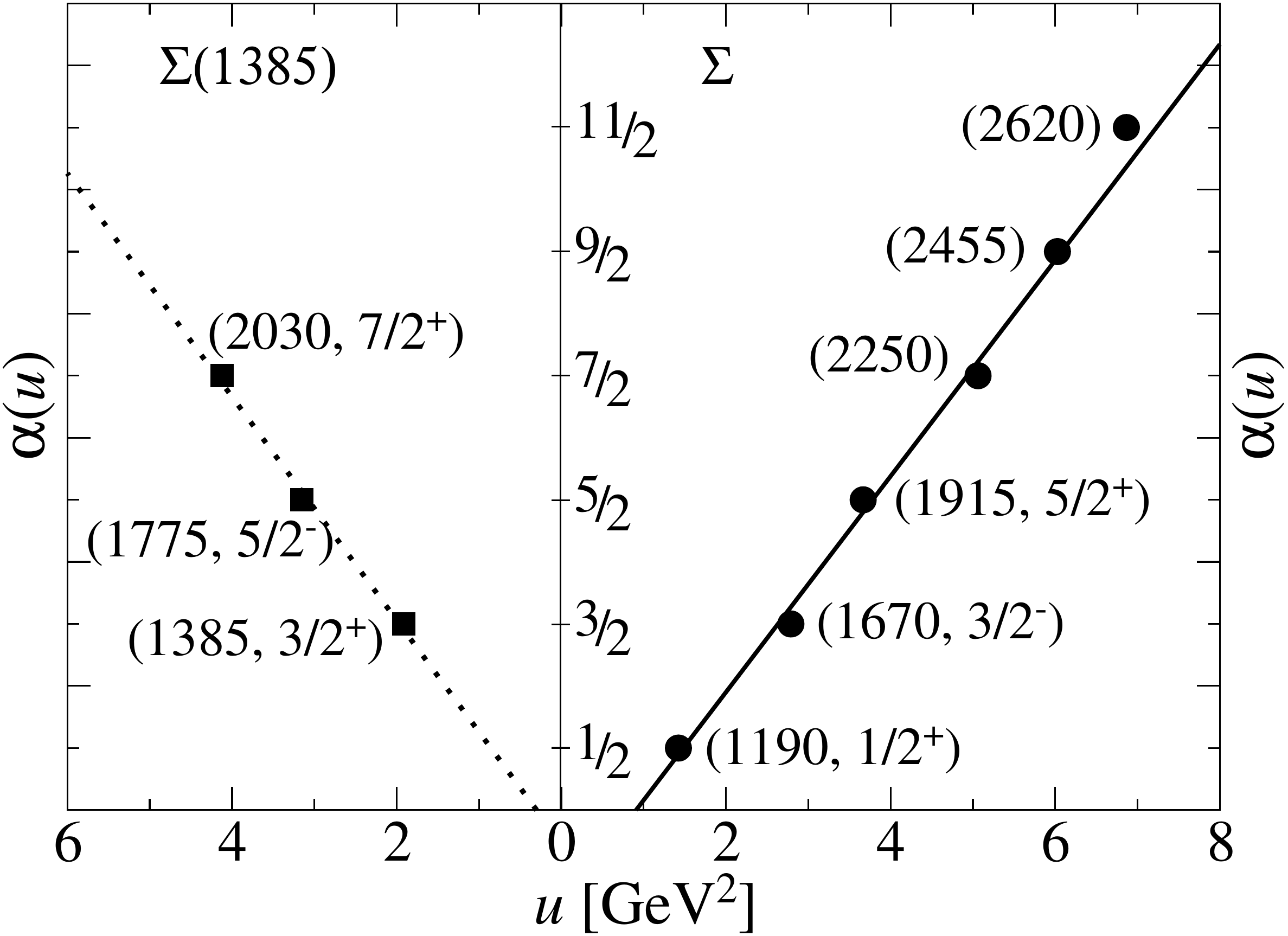}
}
\caption{Regge trajectories of (a) $\Lambda$, and (b) $\Sigma$ and $\Sigma(1385)$,
where the spin and parity of certain hyperon resonances are not confirmed
thus far~\cite{Storrow:1983ct}.}
\label{FIG02}
\end{figure*}

We consider $S=-1$ hyperon Regge trajectories in the $u$ channel that exhibit backward peaks.
The linear Regge trajectories of $\Lambda$, $\Sigma$, and $\Sigma(1385)$ are provided
in~\cite{Storrow:1983ct} with:
\begin{align}
\alpha_\Lambda (u) = -0.65 + 0.94 u,\nonumber\\
\alpha_\Sigma (u) = -0.79 + 0.87 u,\nonumber \\
\alpha_{\Sigma(1385)} (u) = -0.27 + 0.9 u,\nonumber
\end{align}
as shown in Fig.~\ref{FIG02}.

Note, the spin-parities of $\Lambda(2585)$, $\Sigma(2250)$, $\Sigma(2455)$, and
$\Sigma(2620)$ are not experimentally confirmed thus far and are assumed to be $J^P= 11/2^-$, $7/2^-$,
$9/2^+$, and $11/2^-$ in this study, respectively.
The Reggeized $\Lambda$ exchange in the $u$ channel is present only in the $K^- p \to K^+ \Xi^-$
reaction owing to charge conservation.

The $u$-channel Reggeized amplitudes corresponding to the exchanges of $Y(=\Lambda, \Sigma)$ and
$\Sigma^\ast(=\Sigma(1385))$ are obtained as follows:
\begin{eqnarray}
\hskip-0.5cm T_Y (s,u) &=& C_Y(u) \mathcal M_Y^u
\left( \frac{s}{s_Y} \right)^{\alpha_Y(u)-\frac{1}{2}}
\Gamma \left( \frac{1}{2}-\alpha_Y(u) \right) \alpha_Y',  \nonumber \\
\hskip-0.5cm T_{\Sigma^\ast} (s,u) &=& C_{\Sigma^\ast} (u) \mathcal M_{\Sigma^\ast}^u
\left(\frac{s}{s_{\Sigma^\ast}} \right)^{\alpha_{\Sigma^\ast}(u)-\frac{3}{2}}
\hskip-0.2cm\Gamma\left( \frac{3}{2}-\alpha_{\Sigma^\ast}(u) \right) \alpha_{\Sigma^\ast}^\prime,
\label{eq:ReggeAmpl:u-ch}
\end{eqnarray}
by substituting the Regge propagators for the Feynman propagators \cite{Storrow:1983ct}.
The energy-scale parameters were obtained as conventional values of
$s_{Y,\Sigma^\ast} =  1\,\rm{GeV}^2$.
The scale factors were defined as follows ~\cite{Titov:2008yf}:
\begin{align}
C_H (u) = \left(\eta_H \frac{\Lambda_H^2}{\Lambda_H^2 - u} \right)^2,
\label{eq:ScaleFactor}
\end{align}
where $H=(\Lambda, \Sigma, \Sigma(1385))$. Given $\Lambda_H= 1\,\rm{GeV}$, the values of 
$\eta_\Lambda = 2.6$ and $\eta_{\Sigma,\Sigma^\ast} = 0.66$ 
were deduced from the fit of the experimental data in the high-energy region. 

The amplitudes $\mathcal M_{Y,\Sigma^\ast}^u$ in
Eq.~(\ref{eq:ReggeAmpl:u-ch}) are derived from the effective
Lagrangians as follows:
\begin{eqnarray}
{\mathcal M_Y^u} &=& I_Y^u \frac{g_{KNY}}{m_N+m_Y} \frac{g_{K\Xi Y}}{m_\Xi+m_Y} {\bar u}_\Xi
\slashed{k}_1 \gamma_5 (\slashed{q}_u +m_Y) \slashed{k}_2 \gamma_5 u_N,           \nonumber \\
{\mathcal M_{\Sigma^\ast}^u} &=& I_{\Sigma^\ast}^u \frac{g_{KN\Sigma^\ast} g_{K\Xi\Sigma^\ast}}{m_K^2}
{\bar u}_\Xi (\slashed{q}_u+m_{\Sigma^\ast}) \nonumber\\
&{}& \cdot k_1^\mu \Delta_{\mu\nu} (q_u,m_{\Sigma^\ast}) k_2^\nu u_N,
\label{eq:Ampl:u-ch}
\end{eqnarray}
where $u_N$ and $u_\Xi$ denote the spinors of the incoming nucleon and outgoing
$\Xi$ hyperon, respectively.
The baryon Dirac spinors are normalized to $\bar u_B u_B = 1$.
$I_{Y,\Sigma^\ast}^u$ denotes the isospin factor, and the momentum transfer is given by $q_u = p_2 - k_1$.

The PS ($\lambda=1$) and PV ($\lambda=0$) couplings exhibit different behaviors in
the total cross section near the threshold. 
The PS coupling exhibits a sharp increase, whereas the PV coupling exhibits
a slow rise in the cross section. Because the PV coupling fits the experimental data better, we
adopted the PV coupling for ${\mathcal M_Y^u}$ in Eq.~(\ref{Lagrangians:1/2}). 
The element $\Delta_{\mu\nu}$ in the propagator of the spin-3/2 Rarita-Schwinger
field is expressed as follows:
\begin{eqnarray}
\Delta_{\mu\nu} (q,m) &=& -g_{\mu\nu} + \frac13 \gamma_\mu \gamma_\nu  + \frac{1}{3m}
(\gamma_\mu q_\nu - q_\mu \gamma_\nu) \nonumber \\
 &+& \frac{2}{3m^2} q_\mu q_\nu. 
\end{eqnarray}
The coupling constants in Eq.~(\ref{eq:Ampl:u-ch}) can be obtained by using the SU(3)
flavor symmetry relations as follows:
\begin{eqnarray}
g_{KN\Lambda} &=& -13.24,    ~~~~~~~~g_{K\Xi\Lambda} = 3.52,   \nonumber \\
g_{KN\Sigma} &=& 3.58,        ~~~~~~~~~~~~~g_{K\Xi\Sigma} = -13.26,   \nonumber\\
g_{KN\Sigma(1385)} &=& -3.22, ~~g_{K\Xi\Sigma(1385)} = -3.22,  
\end{eqnarray}
with $f/d = 0.575$, $g_{\pi NN} = 13.26$, and $g_{\pi N \Delta} =2.23$~\cite{Nakayama:2006ty}.
 
The differential cross section $d\sigma/du$ satisfies the following asymptotic behavior:
\begin{align}
\frac{d\sigma}{du}(s \to \infty, u \to 0) \propto s^{2\alpha(u)-2}.
\label{eq:Asym:dsdu}
\end{align}

\subsection{Hyperon resonance exchange in the $s$ channel}

We consider the ground and excited states of $\Lambda$ and $\Sigma$ in the $s$ channel, as shown 
in Fig.~\ref{FIG01}(b). Table~\ref{TAB01} lists the properties of 
the (a) $\Lambda$ and (b) $\Sigma$ resonances,
including the decay widths and branching ratios $(\mathcal{B})$ for the $\overline{K} N$ and $K \Xi$
channels from the PDG~\cite{PDG:2022pth} and Ref.~\cite{Sarantsev:2019xxm}.
The branching ratios for $Y^\ast\to K\Xi$ are not well known relative to
those for $Y^\ast\to\overline{K}N$, where $Y^\ast$ denotes $\Lambda$ or $\Sigma$ resonances.
In this study, we include hyperon resonances with known branching ratios and
well-established four-star resonances, such as $\Lambda(1890)3/2^+$, $\Lambda(2100)7/2^-$,
and $\Sigma(2030)7/2^+$. Additionally, we added $\Sigma(2230)3/2^+$ and $\Sigma(2250)$
states to explain the bump structure in the total cross sections near $\sqrt{s}\approx 2.3$ GeV.
Moreover, the spin-parity parameters of $\Sigma(2250)$ with $J^p=7/2^-$ favor the
$\Sigma$ Regge trajectory~\cite{Storrow:1983ct,Bricman:1970ij}.
%

\begin{table}[ht]
\begin{tabular}{cccccc}
\hline\hline
\hspace{-1.3em} $\Lambda$ state&\,Width\,&\,Rating\,
&$\mathcal{B}(\Lambda^\ast\hskip-0.1cm\to\hskip-0.1cm\overline{K} N)$
&$\mathcal{B}(\Lambda^\ast\hskip-0.1cm\to\hskip-0.1cm K \Xi)$
&This \\
$(J^P)$&[MeV]&   &[\%]&[\%]&work \vspace{0.24cm}
\\ \hline
$\Lambda(1820)5/2^+$&80 &****&$55-65$&$-$&   \\
$\Lambda(1830)5/2^-$&90 &****&$4-8$&$-$&   \\
$\Lambda(1890)3/2^+$&120&****&$24-36$&$\sim 1$~\cite{Sarantsev:2019xxm}&\checkmark   \\
$\Lambda(2000)1/2^-$&190&*   &$27 \pm 6$&$-$&   \\
$\Lambda(2050)3/2^-$&493&*   &$19 \pm 4$&$-$&   \\
$\Lambda(2070)3/2^+$&370&*   &$12 \pm 5$&$7 \pm 3$&   \\
$\Lambda(2080)5/2^-$&181&*   &$11 \pm 3$&$4 \pm 1$&   \\
$\Lambda(2085)7/2^+$&200&**  &$-$&$-$&   \\
$\Lambda(2100)7/2^-$&200&****&$25-35$&$< 3$&\checkmark   \\
$\Lambda(2110)5/2^+$&250&*** &$5-25$&$-$&   \\
$\Lambda(2325)3/2^-$&168&*   &$-$&$-$&   \\
$\Lambda(2350)9/2^+$&150&*** &$\sim 12$&$-$& \\
$\Lambda(2585)\hspace{1.1em} ?^? $&   &**  &$-$&$-$& \\
%
%
\hline
\hspace{-1.3em} $\Sigma$ state&\,Width\,&\,Rating\,
&$\mathcal{B}(\Sigma^\ast\hskip-0.1cm\to\hskip-0.1cm\overline{K} N)$
&$\mathcal{B}(\Sigma^\ast\hskip-0.1cm\to\hskip-0.1cm K\Xi)$
&This \\
$(J^P)$&[MeV]&   &[\%]&[\%]&work
\\ \hline
$\Sigma(1880)1/2^+$&200&**  &$10-30$&     $-$&   \\
$\Sigma(1900)1/2^-$&165&**  &$40-70$&     $3 \pm 2$&   \\
$\Sigma(1910)3/2^-$&220&*** &$1-5$&       $-$&   \\
$\Sigma(1915)5/2^+$&120&****&$5-15$&      $-$&   \\
$\Sigma(1940)3/2^+$&250&*   &$13 \pm 2$&  $-$&   \\
$\Sigma(2010)3/2^-$&178&*   &$7 \pm 3$&   $3 \pm 2$&   \\
$\Sigma(2030)7/2^+$&180&****&$17-23$&     $< 2$&\checkmark   \\
$\Sigma(2070)5/2^+$&200&*   &$-$&         $-$&   \\
$\Sigma(2080)3/2^+$&170&*   &$-$&         $-$&   \\
$\Sigma(2100)7/2^-$&260&*   &$8 \pm 2$&   $-$&   \\
$\Sigma(2110)1/2^-$&313&*   &$29 \pm 7$&  $-$&   \\
$\Sigma(2230)3/2^+$&345&*   &$6 \pm 2$&   $2 \pm 1$&\checkmark   \\
$\Sigma(2250)\hspace{1.1em} ?^?$&100&**  &$< 10$&$-$&\checkmark   \\
$\Sigma(2455)\hspace{1.1em} ?^?$&120&*   &$-$&   $-$&   \\
$\Sigma(2620)\hspace{1.1em} ?^?$&200&*   &$-$&   $-$&
\\ \hline\hline
\end{tabular}
\caption{$\Lambda$ and $\Sigma$ resonances listed in the PDG~\cite{PDG:2022pth} with
the decay widths and branching ratios for the $\overline{K} N$ and $K \Xi$ channels.
The resonances with a mark $\checkmark$ are selected in the $s$-channel diagram.}
\label{TAB01}
\end{table}

The scattering amplitude for the exchange of $\Lambda$ and $\Sigma$ in the
$s$ channel can be expressed as follows:
\begin{eqnarray}
{\mathcal M_Y^s} (s) &=& \frac{I_Y^s}{s-m_Y^2}
\frac{g_{KNY}}{m_N+m_Y} \frac{g_{K\Xi Y}}{m_\Xi+m_Y} \nonumber \\
&\cdot&
{\bar u}_\Xi
\slashed{k}_2 \gamma_5 (\slashed{q}_s +m_Y) \slashed{k}_1 \gamma_5 u_N,
\label{eq:Ampl:s-ch}
\end{eqnarray}
with PV coupling in Eq.~(\ref{Lagrangians:1/2}).
The sum of the initial-state momenta is $q_s = k_1 + p_1$.
We consider the hadronic form factor at each vertex in the following form:
\begin{align}
F_Y (s) = \left( \frac{n\Lambda_Y^4}{n\Lambda_Y^4 + (s - m_Y^2)^2} \right)^n,
\end{align}
where the model parameters are fixed at $n=2$ and $\Lambda_ {\Lambda,\Sigma} = 0.85$ GeV.

The scattering amplitudes for the exchange of the $\Lambda$ and $\Sigma$ resonances with
$J^P = 1/2^\pm,\,3/2^\pm,\,5/2^\pm$, and $7/2^\pm$ in the $s$ channel is given by 
the following:
\begin{eqnarray}
\overline{\mathcal M}_{Y^\ast}^{1/2^\pm} &=& -
\Gamma^\pm (\slashed{q}_s + m_{Y^\ast}) \Gamma^\pm,                                      \nonumber\\
\overline{\mathcal M}_{Y^\ast}^{3/2^\pm} &=&
\frac{1}{m_K^2}
\Gamma^\mp (\slashed{q}_s + m_{Y^\ast}) k_2^\mu \Delta_{\mu\nu} k_1^\nu \Gamma^\mp ,         \nonumber\\
\overline{\mathcal M}_{Y^\ast} ^{5/2^\pm} &=&
\frac{-1}{m_K^4}
\Gamma^\pm (\slashed{q}_s + m_{Y^\ast}) k_2^\mu k_2^\nu \Delta_{\mu\nu}^{\alpha\beta}
k_{1\alpha} k_ {1\beta} \Gamma^\pm  ,                                                   \nonumber\\
\overline{\mathcal M}_{Y^\ast}^{7/2^\pm} &=&
\frac{1}{m_K^6}
\Gamma^\mp (\slashed{q}_s + m_{Y^\ast}) k_2^\mu k_2^\nu k_2^\rho \Delta_{\mu\nu\rho}^{\alpha\beta\delta}
k_{1\alpha} k_{1\beta} k_{1\delta}  \Gamma^\mp, \hskip+0.5cm
\label{Ampl:Resonances}
\end{eqnarray}
with the notation:
\begin{align}
{\mathcal M}_{Y^\ast}^{J^P} (s) = I_{Y^\ast}
\frac{g_{KNY^\ast} g_{K\Xi Y^\ast}}{s-M_{Y^\ast}^2 + i m_{Y^\ast} \Gamma_{Y^\ast}}
\bar u_\Xi \overline{\mathcal M}_{Y^\ast}^{J^P} u_N,
\end{align}
where $\Gamma^+ \equiv \gamma_5$ and $\Gamma^- \equiv 1_{4 \times 4}$.
We refer to Refs.~\cite{Behrends:1957rup,Rushbrooke:1966zz,Chang:1967zzc,Man:2011np} for
the elements $\Delta_{\mu\nu(\rho)}^{\alpha\beta(\delta)}$ involved in the propagators of the
spin-$5/2$ and $7/2$ baryon fields.
The decay widths $\Gamma_{Y^\ast}$ were obtained from the PDG~\cite{PDG:2022pth} shown in
Table~\ref{TAB01}.
The mass of $\Sigma(2250)$ is assumed to be 2290 MeV in this study to fit the bump
structure near $\sqrt{s} \approx 2.3$ GeV.
The Gaussian form factor is considered at each vertex such that the resonance amplitudes
vanish at high energies:
\begin{eqnarray}
F_{Y^\ast}(s) = \exp\left(-\frac{s-m_{Y^\ast}^2}{\Lambda_{Y^\ast}^2} \right),
\end{eqnarray}
where the cutoff energy is chosen to be $\Lambda_{Y^\ast} = 0.73$ GeV.

The coupling constants $g_{KNY^\ast}$ are determined from the branching ratios of
$Y^\ast\to \overline{K} N$.
We adopt the central values for $\mathcal{B}(Y^\ast\to\overline{K}N)$.
For the $\Sigma(2250)\to\overline{K}N$ decay, only the upper limit of the branching ratio is
available, thus we obtain the upper-limit value \cite{PDG:2022pth}.
We then obtained the values of $g_{K \Xi Y^\ast}$ and $\mathcal{B}(Y^\ast\to K\Xi)$ by fitting the
experimental data.
All relevant values are listed in Table~\ref{TAB02}.
The extracted values of $\mathcal{B}(Y^\ast\to K\Xi)$ sufficiently agreed with
those quoted in the PDG~\cite{PDG:2022pth}.
The branching ratio of the $\Sigma(2250)\to K\Xi$ decay was 1.0 \%.
\begin{table}[ht]
\begin{tabular}{cccccc}
\hline\hline
&$\Lambda(1890)$ & $\Lambda(2100)$ & $\Sigma(2030)$ & $\Sigma(2230)$ & $\Sigma(2250)$
\\ \hline
$g_{K N Y^\ast}$& 0.84 & 2.41 & 0.82 & 0.41 & 0.59 \\
$g_{K \Xi Y^\ast }$& $-0.26$ & 2.95 & $-0.93$ & 0.34 & 0.88 \\
$\mathcal{B}(Y^\ast \to K \Xi)[\%]$ & 0.23 & 0.73 & 0.88 & 2.0 & 1.0 \\
\hline\hline
\end{tabular}
\caption{
The coupling constants $g_{K N Y^\ast}$ were determined from the central values of
$\mathcal{B}(Y^\ast\to \overline{K} N$)~\cite{PDG:2022pth}.
The other values of $g_{K\Xi Y^\ast}$ and $\mathcal{B}(Y^\ast \to K \Xi)$ were extracted by fitting
the experimental data for the $K^- p \to K \Xi$ reactions.
}
\label{TAB02}
\end{table}

\subsection{Rescattering diagram}

The meson-baryon rescattering amplitude can be expressed as follows:
\begin{eqnarray}
\hskip-0.4cm T_{MB}(p,p') &=& \sum_i \int
\frac{d^3 \mathbf{q}}{(2\pi)^3}
\frac{m_{B_i}}{E_{B_i}} T_{K^- p \to M_i B_i}(p,q) \nonumber\\
&\cdot&
\frac{1}{s - (E_{M_i} + E_{B_i})^2 + i \epsilon}
T_{M_i B_i \to K \Xi}(q,p'), 
\end{eqnarray}
which is derived from the three-dimensional reduction of the Bethe-Salpeter equation.
The off-shell energies of the intermediate meson and baryon are denoted by
$E_{M_i} = (m_{M_i}^2+|\mathbf{q}|^2)^{1/2}$ and $E_{B_i} = (m_{B_i}^2+|\mathbf{q}|^2)^{1/2}$,
respectively.
For the intermediate states, we consider $M_i = (\rho, \omega, \phi, \pi, \eta)$ and
$B_i = (\Lambda, \Sigma)$, as shown in Fig.~\ref{FIG01}(c) and (d).
The summation over the polarization and spin indices of the intermediate states is
indicated.
For both the production $T_{K^- p \to M_i B_i}$ and absorption $T_{M_i B_i \to K \Xi}$ processes,
the $K^\ast$ exchange is considered.
In the case of $M_i = (\rho, \omega, \phi)$, we also include the $K$ exchange.

The integral runs over the three-momentum $\mathbf{q}$ of the intermediate $M_i$ meson and can
be decomposed into singular and principal parts $\mathcal{P}$:
\begin{eqnarray}
T_{MB}(p,p') &=& -i \sum_i \frac{q_{\rm{c.m.}}}{16\pi^2} \frac{m_{B_i}}{\sqrt s}
\int d \Omega 
\nonumber \\
&{}&\left[ T_{K^- p \to M_i B_i}(p,q)
T_{M_i B_i \to K\Xi}(q,p') \right] + \mathcal{P},\hskip+0.6cm
\end{eqnarray}
where the former is strictly required by unitarity.
In this study, we numerically consider both parts
by using a well-established method.
$q_{\rm{c.m.}} = [(s-(m_{M_i}-m_{B_i})^2)(s-(m_{M_i}+m_{B_i})^2)/4s]^{1/2}$ denotes the magnitude
of the on-shell three-momentum of the intermediate hadrons.

The transition amplitude $ T_{K^- p \to M B}$ can be constructed by using the following
effective Lagrangians:
\begin{eqnarray}
{\mathcal L}_{VKK } &=& -i g_{VKK}(\partial^\mu \overline{K} V_\mu K - \overline{K} V_\mu \partial^\mu K),\nonumber\\
{\mathcal L}_{VK^ \ast K} &=& g_{VK^\ast K} \epsilon^{\mu\nu\alpha\beta}
\partial_\mu \overline{K}^\ast_\nu \partial_\alpha V_\beta K + \rm{H.c.},                     \nonumber\\
{\mathcal L}_{K^\ast KP} &=& -i g_{K^\ast KP} (\overline{K}\partial^\mu P K^\ast_\mu -
\overline{K}^\ast_\mu \partial^\mu P K),                                                    \nonumber\\
{\mathcal L}_{K^\ast NB} &=& - g_{K^\ast NB} \left[ \overline{K}^\ast_\mu \overline{B} \gamma^\mu -
\frac{\kappa_{K^\ast NB}}{2M_N} \partial_\nu \overline{K}^\ast_\mu \overline{B} \sigma^{\mu \nu}\right] N \nonumber\\
&+& \rm{H.c.}, \hskip+0.5cm
\label{Lagrangians}
\end{eqnarray}
where $V=(\phi,\rho,\omega)$, $P=(\pi,\eta)$, and $B=(\Lambda,\Sigma)$.
The coupling constant $g_{\phi KK}=4.48$ was calculated from the partial decay width as follows:
$\Gamma_{\phi\to K^+ K^-}=2.09$ MeV~\cite{PDG:2022pth}.
\begin{align}
\Gamma_{\phi \to K^+ K^-} = \frac{g_{\phi KK}^2 q_K^3}{6\pi m_\phi^2},
\label{DW:Phi_KK}
\end{align}
where $q_K=(m_\phi^2-4m_K^2)^{1/2}/2$.
The SU(3) flavor symmetry leads to~\cite{Oh:2004wp}:
\begin{align}
& g_{\rho KK} = g_{\omega KK} =g_{\rho\pi\pi}/2,                                             \cr
& g_{\phi K^\ast K} = g_{\rho\omega\pi}/\sqrt2,\,\,\,
  g_{\rho K^\ast K} = g_{\omega K^\ast K} = g_{\rho\omega\pi}/2,                                     \cr
& g_{K^\ast K \eta} = \sqrt3 g_{K^\ast K \pi},
\label{DW:RhoOmega_KK}
\end{align}
where $g_{\rho\pi\pi} = 5.94$ was calculated from the branching ratio $\mathcal{B}(\rho \to \pi \pi)
\sim 1$. The $g_{\rho\omega\pi}$ coupling constant was derived from the hidden gauge
approach~\cite{Bando:1987br}:
\begin{align}
g_{\rho\omega\pi} = \frac{N_c g_{\rho\pi\pi}^2}{8 \pi^2 f_\pi} = 14.4 \, \rm{GeV^{-1}},
\label{DW:RhoOmega_KK}
\end{align}
where $N_c$ = 3 and $f_\pi$ = 93 MeV.
We obtained $g_{K^\ast K \pi} = 6.56$ from the branching ratio $\mathcal{B}(K^\ast \to K\pi) \sim 1$ with
the relation:
\begin{align}
\Gamma_{K^\ast \to K \pi} = \frac{g_{K^\ast K \pi}^2 q_\pi^3}{8\pi m_{K^\ast}^2},
\label{DW:Ks_KPi}
\end{align}
where $q_\pi = [(M_{K^\ast}^2-(m_K -m_\pi )^2)(M_{K^\ast}^2-(m_K+m_\pi )^2)]^{1/2}/2M_{K^\ast}$.
The $K^\ast N B$ coupling constants were obtained from the Nijmegen potential
(NSC97a)~\cite{ Stoks:1999bz,Rijken:1998yy}:
\begin{align}
g_{K^\ast N \Lambda} = 4.26,& \,\,\, k_{K^\ast N \Lambda} = 2.66,   \cr
g_{K^\ast N \Sigma} = -2.46,& \,\,\, k_{K^\ast N \Sigma} = -0.467. 
\label{Ampl:Box1}
\end{align}
The $K^\ast \Xi B$ interactions were determined in a similar manner:
\begin{align}
g_{K^\ast \Xi \Lambda} = 4.26,& \,\,\, k_{K^\ast \Xi \Lambda} = 1.10,   \cr
g_{K^\ast \Xi \Sigma} = -2.46,& \,\,\, k_{K^\ast \Xi \Sigma} = 4.22, 
\label{Ampl:Box1}
\end{align}
which are needed for the transition amplitude $T_{M B \to K \Xi}$.

\begin{figure}[ht]
\includegraphics[width=7.5cm]{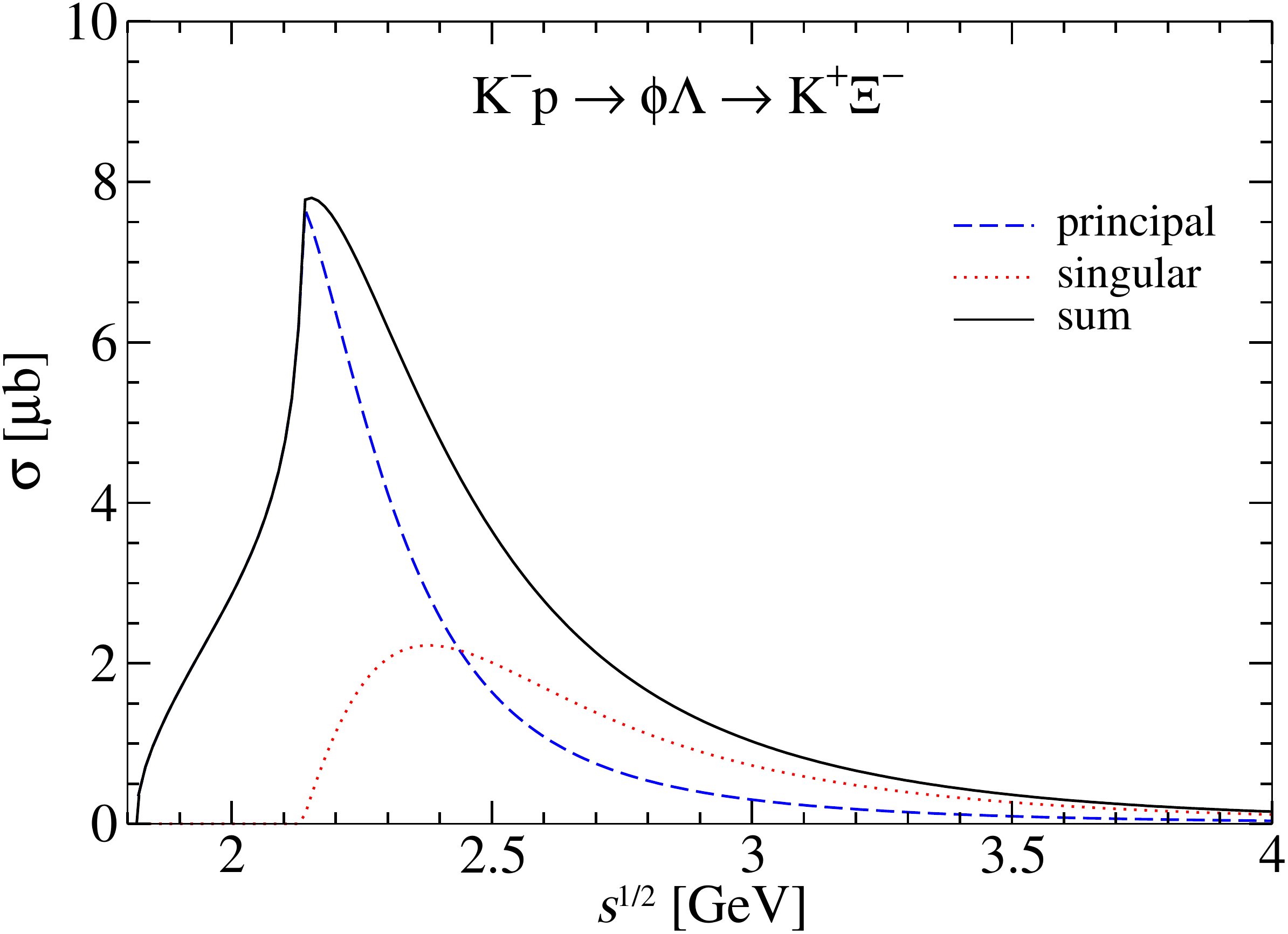}
\caption{Total cross section from the $\phi \Lambda$ rescattering diagram versus $\sqrt{s}$ for
$K^- p \to K^+ \Xi^-$.
The dashed blue and dotted red lines indicate the contributions of the principal and
singular parts, respectively.
The solid black line indicates the sum of both contributions.}
\label{FIG03}
\end{figure}

For the process of $K^-(k_1) + p (p_1) \to M (q') + B (q'')$, the transition amplitudes
in Fig.~\ref{FIG01}(c) can be expressed as follows:
\begin{eqnarray}
T_{K^- p \to V B}^K &=& i
\frac{g_{KNB}}{m_N+m_B} \frac{g_{VKK}}{q_t^2-m_K^2} (k_1 + q_t)^\mu \epsilon^\ast_\mu
\bar u_\Xi \slashed{q}_t \gamma_5 u_N,                                             \nonumber\\
T_{K^- p \to V B}^{K^\ast} &=& \frac{g_{V K^\ast K} g_{K^\ast NB}}{q_t^2-m_{K^\ast}^2} \epsilon^{\mu\nu\alpha\beta}
\bar u_\Xi  \nonumber\\
&\cdot&
\left[ \gamma_\mu + i \frac{\kappa_{K^\ast NB}}{2m_N} \sigma_{\mu\lambda} q_t^\lambda \right] u_N
q_{t\nu} q'_\alpha \epsilon^\ast_\beta,
\label{Ampl:Box1}
\end{eqnarray}
for the $K$ and $K^\ast$ exchanges, respectively.
Here, $q_t = k_1-q'=[E_{K^-}(k_1)-E_M(q'),~\mathbf{p_{K^-}}-\mathbf{q}]$ and ${\epsilon}_\mu$
is the polarization vector of the $V$ meson.
The transition amplitude corresponding to Fig.~\ref{FIG01}(d) has the following form:
\begin{eqnarray}
T_{K^- p \to P B}^{K^\ast} &=& \frac{g_{K^\ast K P} g_{K^\ast N B}}{q_t^2-m_{K^\ast}^2}
q'_\alpha \left[ -g^{\mu\alpha} + \frac{q_t^\mu q_t^\alpha}{m_{K^\ast}^2} \right] \bar u_\Xi \nonumber\\
&\cdot&
\left[ \gamma_\mu + i \frac{\kappa_{K^\ast NB}}{2m_N} \sigma_{\mu\nu} q_t^\nu \right] u_N,
\label{Ampl:Box2}
\end{eqnarray}
for the $K^\ast$ exchange.

We considered the following form factor at each vertex:
\begin{align}
F_{MB} (\mathbf{q}_t) =
\left( \frac{\Lambda_{MB}^2}{\Lambda_{MB}^2 + \mathbf{q}_t^2} \right)^2 ,
\end{align}
where $\mathbf{q}_t^2 = (\mathbf{p}-\mathbf{q})^2$.
The cut-off energies were fixed at $\Lambda_{VB} = 0.85$ GeV and $\Lambda_{PB} = 0.50$ GeV.

The contribution of the $\phi \Lambda$ rescattering diagram for $K^- p \to K^+ \Xi^-$ is
shown in Fig.~\ref{FIG03}.
It was found that the principal part plays an essential role in the low-energy region,
whereas the singular part governs the high-energy region.
Note, the singular part requires only the on-shell amplitudes
for
$T_{K^- p \to \phi \Lambda}$ and $T_{\phi \Lambda \to K^+ \Xi^-}$, 
such that it contributes from
the $\phi \Lambda$ threshold.
For the early theortical attempts to consider the rescattering diagrams 
for the meson-induced reactions in Refs.~\cite{Singh:1969va,Saxena:1970vs,Agarwal:1971kb}, the
angular distributions were explained only by the rescattering diagrams with certain assumptions.

\section{Numerical results} \label{Section:III}

We first present our numerical results for the total and differential cross sections, and
recoil polarization asymmetries using only the $u$- and $s$-channel diagrams.
We then discuss how the rescattering
diagrams became important when they were included.

\begin{figure}[ht]
\stackinset{r}{0.5cm}{t}{0.3cm}{(a)}{
\includegraphics[width=8.5cm]{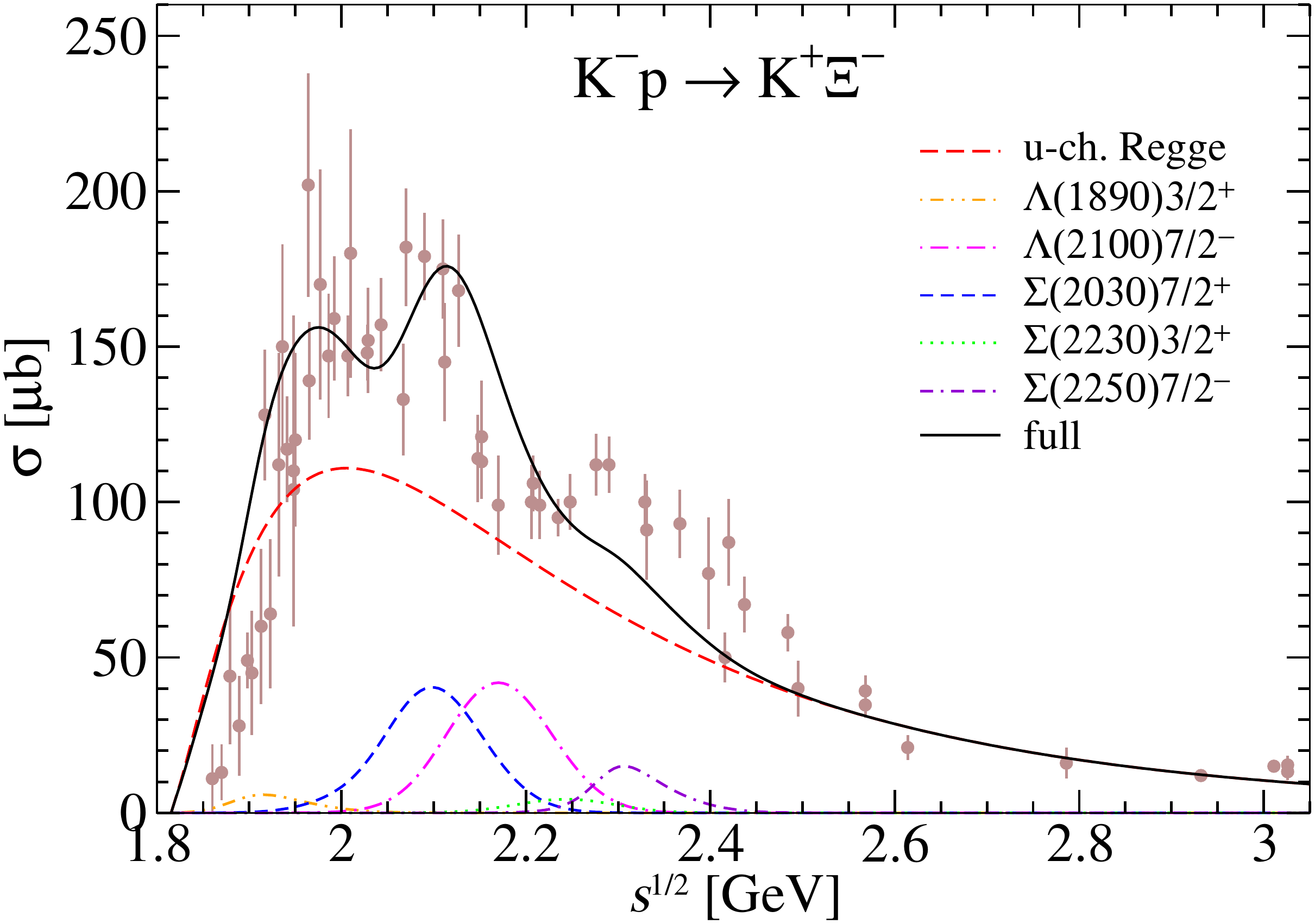}
}\hspace{1em}
\stackinset{r}{0.5cm}{t}{0.3cm}{(b)}{
\includegraphics[width=8.5cm]{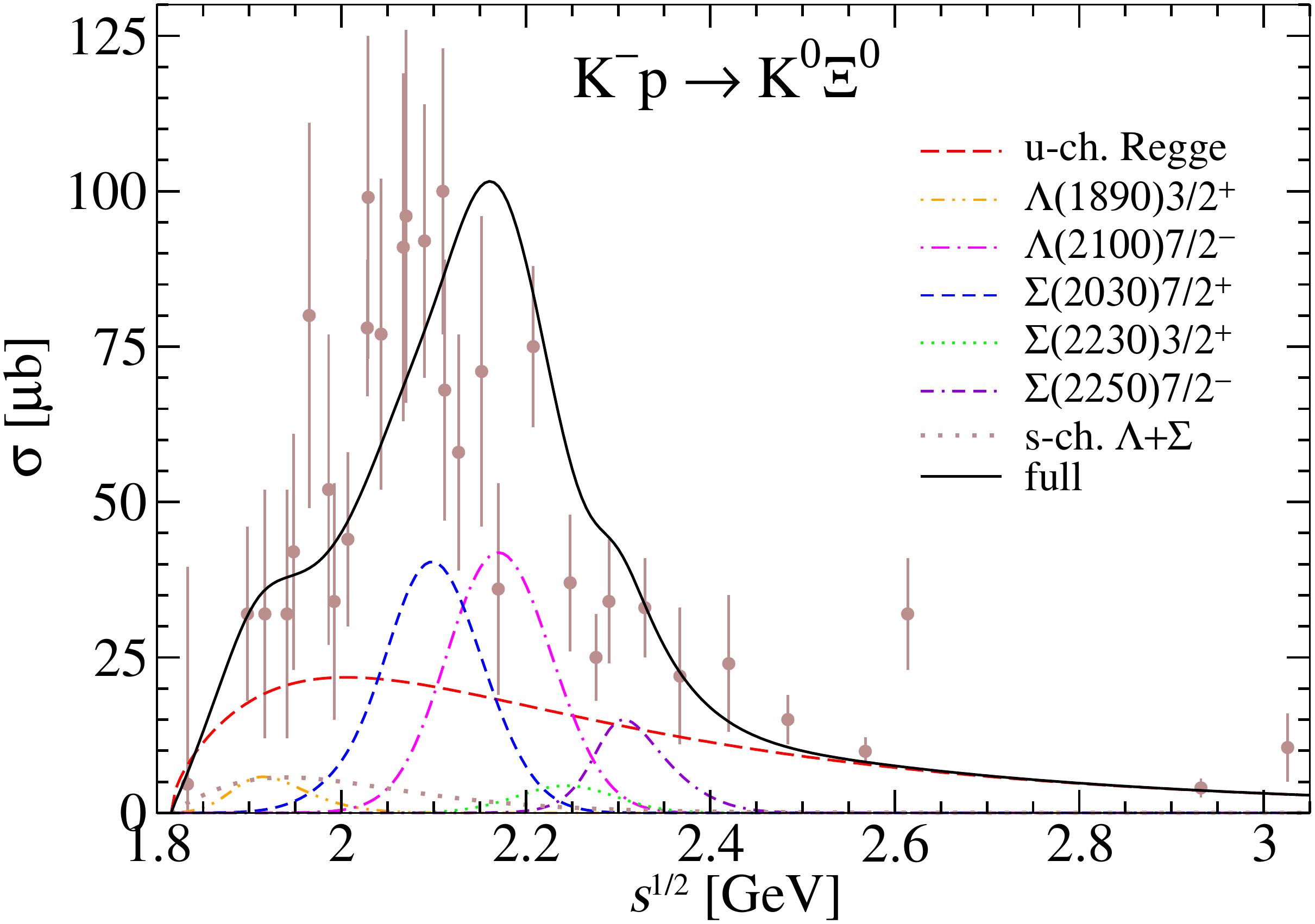}
}
\caption{Total cross sections as a function of the c.m. energy $\sqrt{s}$ for 
the (a) $K^- p \to K^+\Xi^-$ and (b) $K^- p \to K^0 \Xi^0$ reactions.
Contributions from the $s$-channel hyperon resonances involving $\Lambda(1890)3/2^+$,
$\Lambda(2100)7/2^-$, $\Sigma(2030)7/2^+$, $\Sigma(2230)3/2^+$, and $\Sigma(2250)7/2^-$ are shown
separately.
The dashed red line indicates the $u$-channel contributions of the $\Lambda$, $\Sigma$, and
$\Sigma (1385)$ Regge trajectories.
The dotted brown line indicates the contributions of the
$s$-channel $\Lambda$ and $\Sigma$ ground states, and is only observed in the $K^0\Xi^0$ channel.
The solid black line indicates the full contribution of Fig.~\ref{FIG01}(a) and (b).
The data are taken from (a) Refs.~\cite{Carmony:1964zza,Berge:1966zz,Birmingham:1966onr,
London:1966zz, Trippe:1967wat,Trower:1968zz,Burgun:1968ice,Dauber:1969hg,SABRE:1971pzp,
deBellefon:1972rjq,Rader:1973ja,Griselin:1975pa,Briefel:1977bp} and (b)
Refs.~\cite{Berge:1966zz,Dauber:1969hg,SABRE:1971pzp,deBellefon:1972rjq,Briefel:1977bp,
Carlson:1973td}.}
\label{FIG04}
\end{figure}


\subsection{Regge-plus-resonance model }

The result of the total cross section is shown in Fig.~\ref{FIG04} as a function of the
c.m. energy $\sqrt{s}$ with individual contributions for (a) $K^- p \to K^+ \Xi^-$ and (b)
$K^- p \to K^0 \Xi^0$, and compared with the available experimental
data~\cite{Carmony:1964zza,Berge:1966zz,Birmingham:1966onr,London:1966zz, Trippe:1967wat,
Trower:1968zz,Burgun:1968ice,Dauber:1969hg,SABRE:1971pzp,deBellefon:1972rjq,Rader:1973ja,
Griselin:1975pa,Briefel:1977bp,Carlson:1973td}.
For a $u$-channel isospin-1 exchange mechanism, we obtain the relation $\sigma(K^- p \to
K^0 \Xi^0)/\sigma(K^- p \to K^+ \Xi^-) = 4$.
The $\Lambda$ exchange in the $u$ channel is only possible in the $K^+ \Xi^-$ channel.
Thus, the Reggeized $\Sigma$ and $\Sigma(1385)$ exchanges are fitted to describe the high
energy data ($\sqrt{s} > 2.4$ GeV) of $K^- p \to K^0 \Xi^0$, and the background
contribution of $K^- p \to K^+ \Xi^-$ originates mostly from the $u$-channel $\Lambda$ Regge
trajectory.
The coherent sum of the $s$-channel amplitudes involving the $\Lambda$ and $\Sigma$ ground states
makes a significant contribution only to the $K^0\Xi^0$ channel because the $\Lambda$ and
$\Sigma$ channels interfere constructively only in the $K^0\Xi^0$ channel, and not in the $K^+
\Xi^-$ channel.

The bump structures at $1.9< \sqrt{s}< 2.4$ GeV were accounted for by the
additional inclusion of the $s$-channel $\Lambda$ and $\Sigma$ resonances in the
background contribution.
The most significant contributions are from $\Lambda(2100)7/2^-$ and $\Sigma(2030)7/2^+$.
Near the threshold, $\Lambda(1890)3/2^+$, which is the lowest state among the $s$-channel 
resonances in this study, contributes to the description of the
experimental data. However, an apparent discrepancy was observed in the increasing pattern of 
the total cross section for $K^-p\to K^+\Xi^-$  
near the threshold, indicating significant contributions from subthreshold resonances 
\cite{Ikeda:2011dx,CLAS:2017sgi}. 
The $\Sigma(2230)3/2^+$ and $\Sigma(2250)$ resonances contribute moderately.
As previously indicated, we assumed $\Sigma(2250)$ to have $J^P = 7/2^-$, lying on the
Regge trajectory of $\Sigma$ in Fig.~\ref{FIG02}.
The magnitudes of the individual $s$-channel resonance contributions are the same for
both channels.

\begin{figure}[ht]
\stackinset{c}{-1.cm}{t}{0.1cm}{(a)}{
\includegraphics[width=8.5cm]{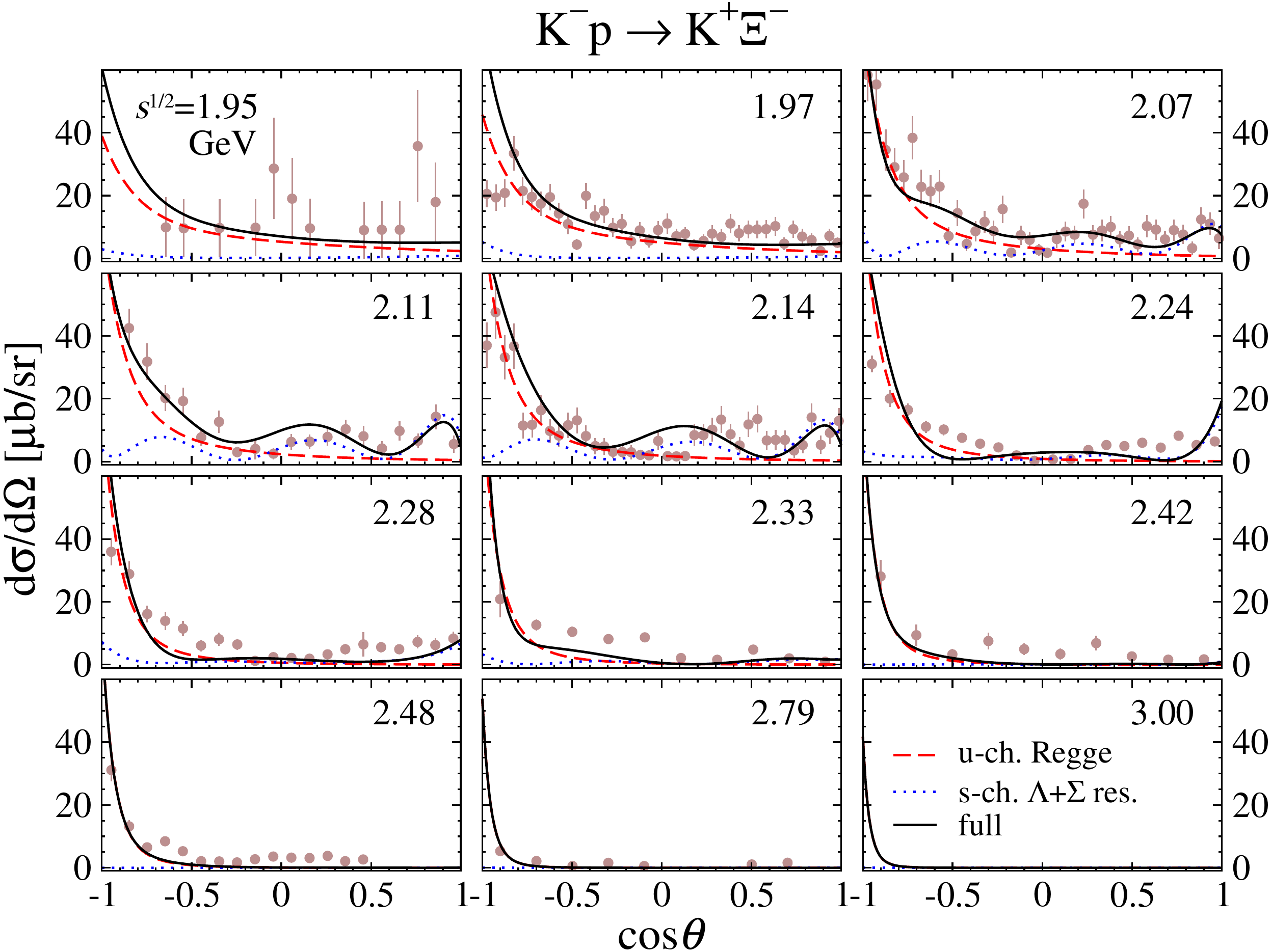}
}\,\,\,
\stackinset{c}{-1.cm}{t}{0.1cm}{(b)}{
\includegraphics[width=8.5cm]{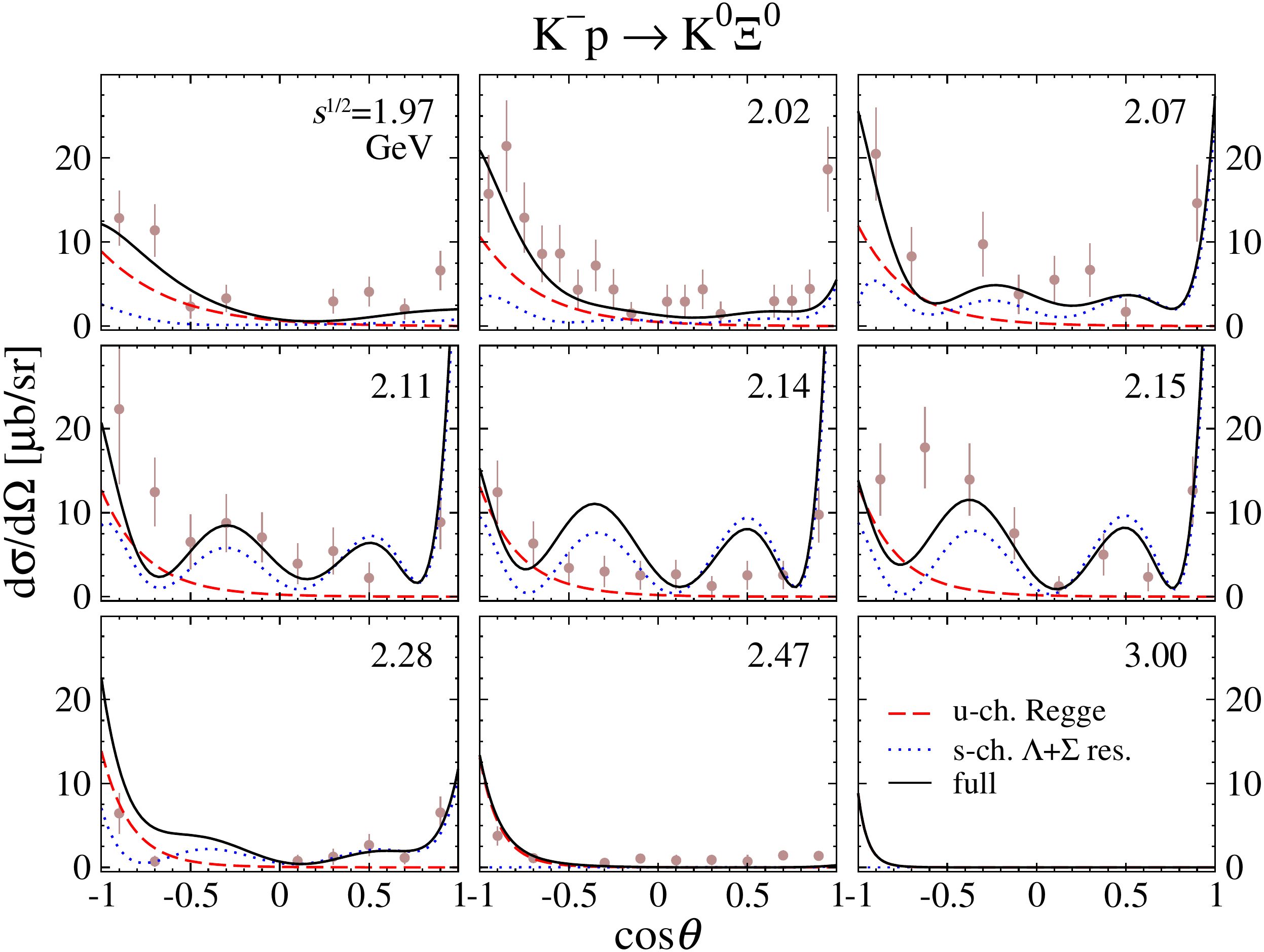}
}
\caption{Differential cross sections versus $\cos\theta$ for (a) $K^- p \to K^+ \Xi^-$ and
(b) $K^- p \to K^0 \Xi^0$ for different c.m. energies $\sqrt{s} = (1.95 $--$ 3.00)$ GeV.
The dashed red lines are the contributions of the $u$-channel $\Lambda$, $\Sigma$, and
$\Sigma (1385)$ Regge trajectories.
The dotted blue lines represent the contributions of the $s$-channel $\Lambda$ and
$\Sigma$ resonances.
The solid black lines indicate the full contribution of Fig.~\ref{FIG01}(a) and (b).
The data are from (a) Refs.~\cite{Birmingham:1966onr,London:1966zz,Trippe:1967wat,
Trower:1968zz,Burgun:1968ice,Dauber:1969hg} and (b) Refs.~\cite{Berge:1966zz,
Burgun:1968ice,Dauber:1969hg,Carlson:1973td}.}
\label{FIG05}
\end{figure}

Figure~\ref{FIG05} presents differential cross sections as a function of
$\cos\theta$ for (a) $K^- p \to K^+ \Xi^-$ and (b) $K^- p \to K^0 \Xi^0$ reactions, where $\theta$ is
the scattering angle of the outgoing kaon in the c.m. frame.
In both channels, the backward-angle scattering is attributed to the exchange of the
$\Lambda$, $\Sigma$, and $\Sigma(1385)$ Regge trajectories in all the energy regions.
However, the two channels exhibited different shapes at the forward angles in the
resonance region $2.07<\sqrt{s}< 2.15$ GeV.
The differential cross sections drastically decrease at significantly forward angles
($\cos\theta > 0.9$) in the $K^+\Xi^-$ channel; however,
sharply increase in the $K^0\Xi^0$ channel.
The two distinct behaviors at the significantly forward angles are due to the varying
interference between the $\Lambda(2100)7/2^-$ and $\Sigma(2030)7/2^+$ amplitudes based on their
isospins.
They interfere destructively in the $K^+\Xi^-$ channel, but interfere constructively in the $K^0\Xi^0$
channel at forward angles.

Note, the $K^- p \to K\Xi^\ast(1530)$ reactions exhibit the opposite behavior at forward angles ~\cite{Dauber:1969hq}.
Sharp forward peaking was observed only in the $K^-p \to K^+\Xi^{*-}$ reaction at
$\sqrt{s} = 2.27$ and $2.43$ GeV, but not in the $K^- p \to K^0 \Xi^{*0}$ reaction.
The lack of information regarding the $Y^\ast \to K \Xi^\ast$ decays~\cite{PDG:2022pth} limits
detailed theoretical analysis of the $K^-p \to K \Xi^{*}$ reaction; therefore,
further studies are warranted.

\begin{figure}[ht]
\stackinset{l}{0.3cm}{t}{0.15cm}{(a)}{
\includegraphics[width=7.5cm]{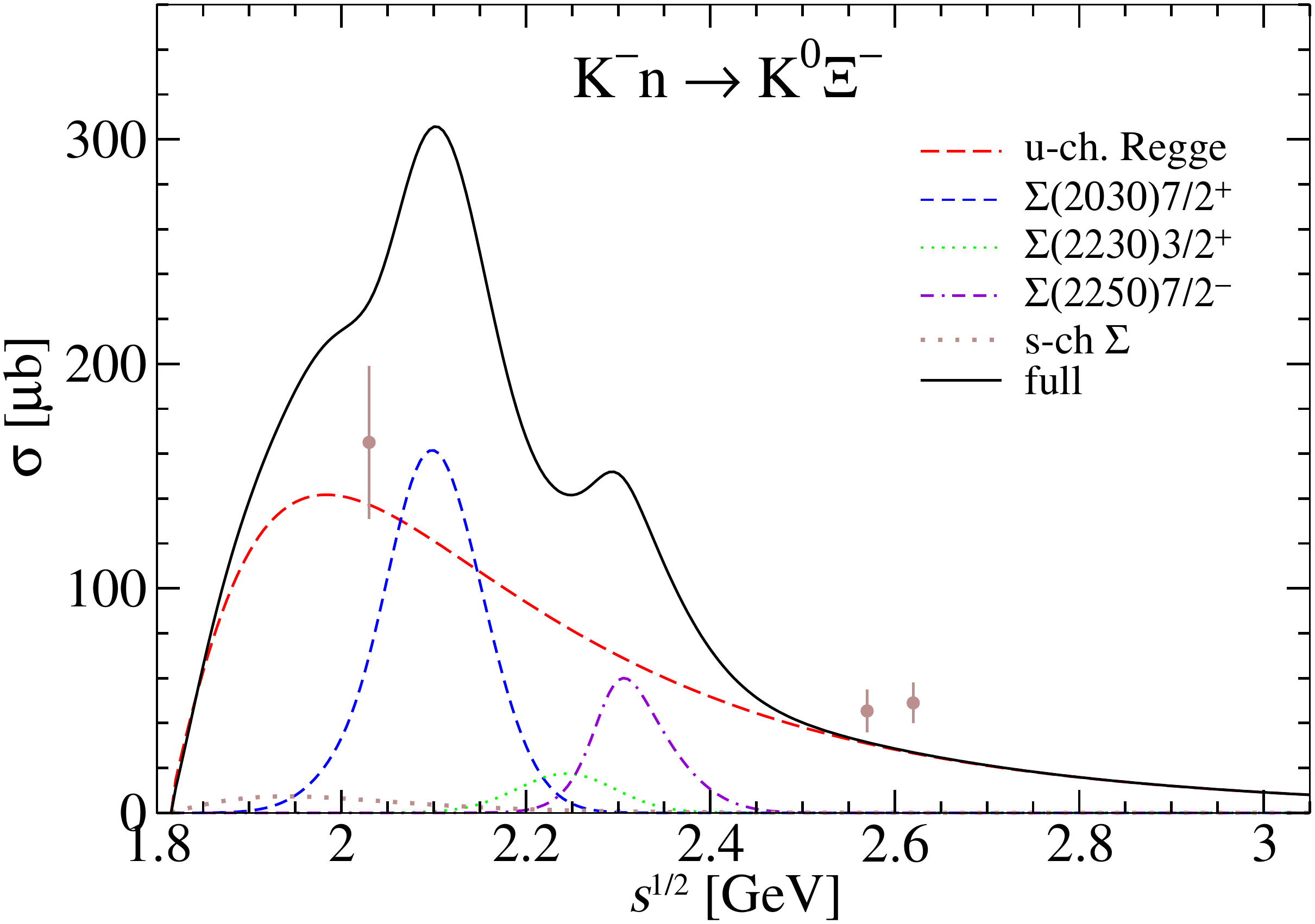}
} \hspace{1em}
\stackinset{l}{0.3cm}{t}{0.15cm}{(b)}{
\includegraphics[width=7.3cm]{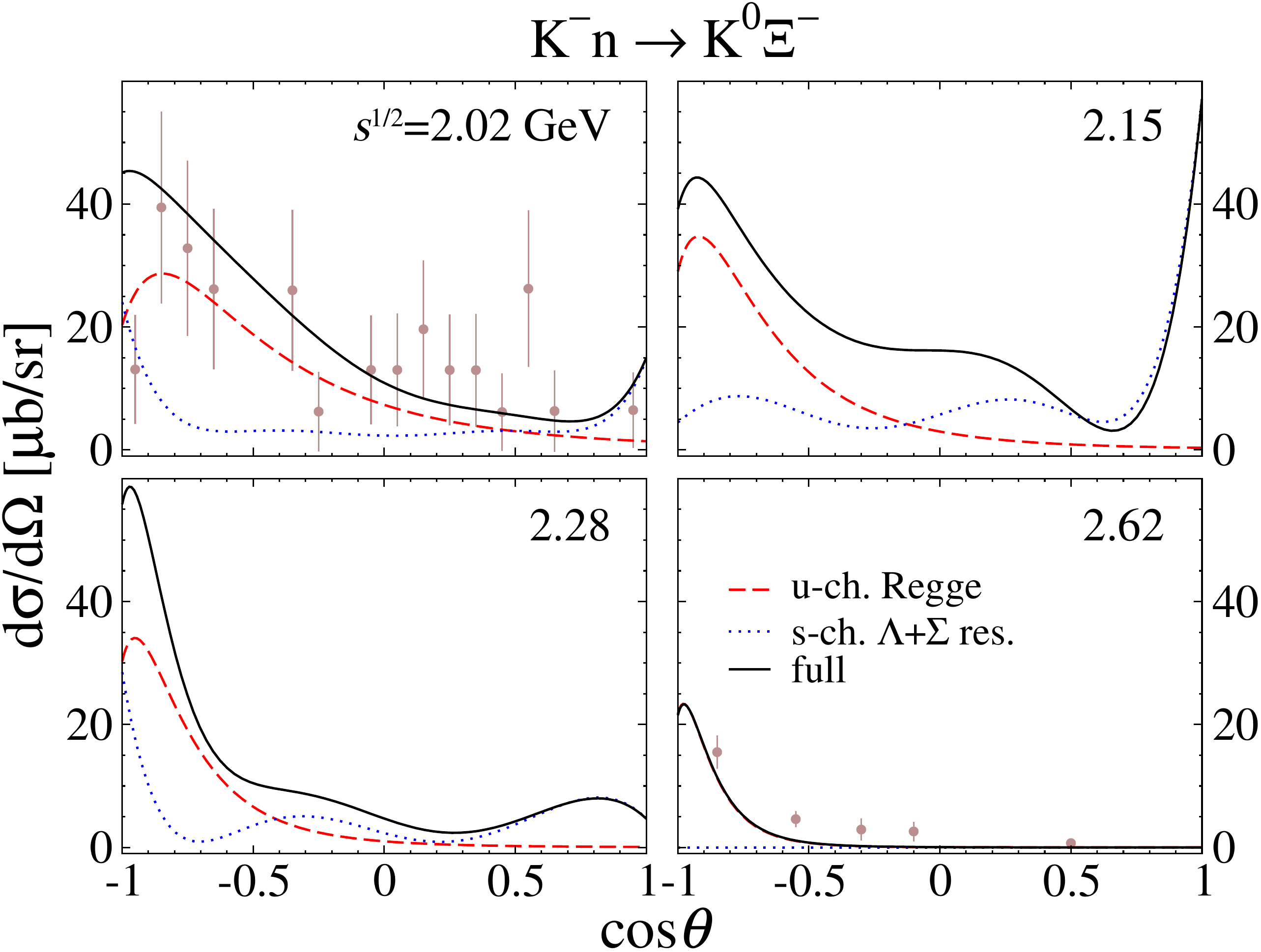}
}
\caption{The same as in (a) Fig.~\ref{FIG04} and (b) Fig.~\ref{FIG05} but for
$K^- n \to K^0 \Xi^-$.
The data are obtained from Refs.~\cite{Berge:1966zz,SABRE:1971pzp}.}
\label{FIG06}
\end{figure}
Our model predictions are extended to the $K^- n \to K^0 \Xi^-$ reaction as shown in
Fig.~\ref{FIG06}.
This provides a good test ground for studying hyperon resonances 
because the $s$-channel $I=0$ exchange is forbidden.
The Reggeized hyperon exchange in the $u$ channel is slightly enhanced compared to the
$K^- p \to K^+ \Xi^-$ reaction shown in Fig.~\ref{FIG04}(a) because the interference between
the Reggeized $I=0$ $\Lambda$ and $I=1$ $\Sigma$ exchanges change.
The $\Sigma(2030)7/2^+$ and $\Sigma(2250)$ contribute most strongly to the $s$-channel resonances
in the $\sqrt{s}\approx 2.1$ and $\sqrt{s} \approx 2.3$ GeV regions, respectively.
The differential cross sections at $\sqrt{s} = 2.02$ GeV
decrease monotonically as $\cos\theta$ increases from $-1.0$ to $0.8$.
The forward peak at $\sqrt{s} = 2.15$ GeV disappears at $2.28$ GeV in Fig.~\ref{FIG06}(b)
owing to the different resonances involved.
The calculated results were reconsiled with the experimental data.

\begin{figure}[htb]
\includegraphics[width=8.5cm]{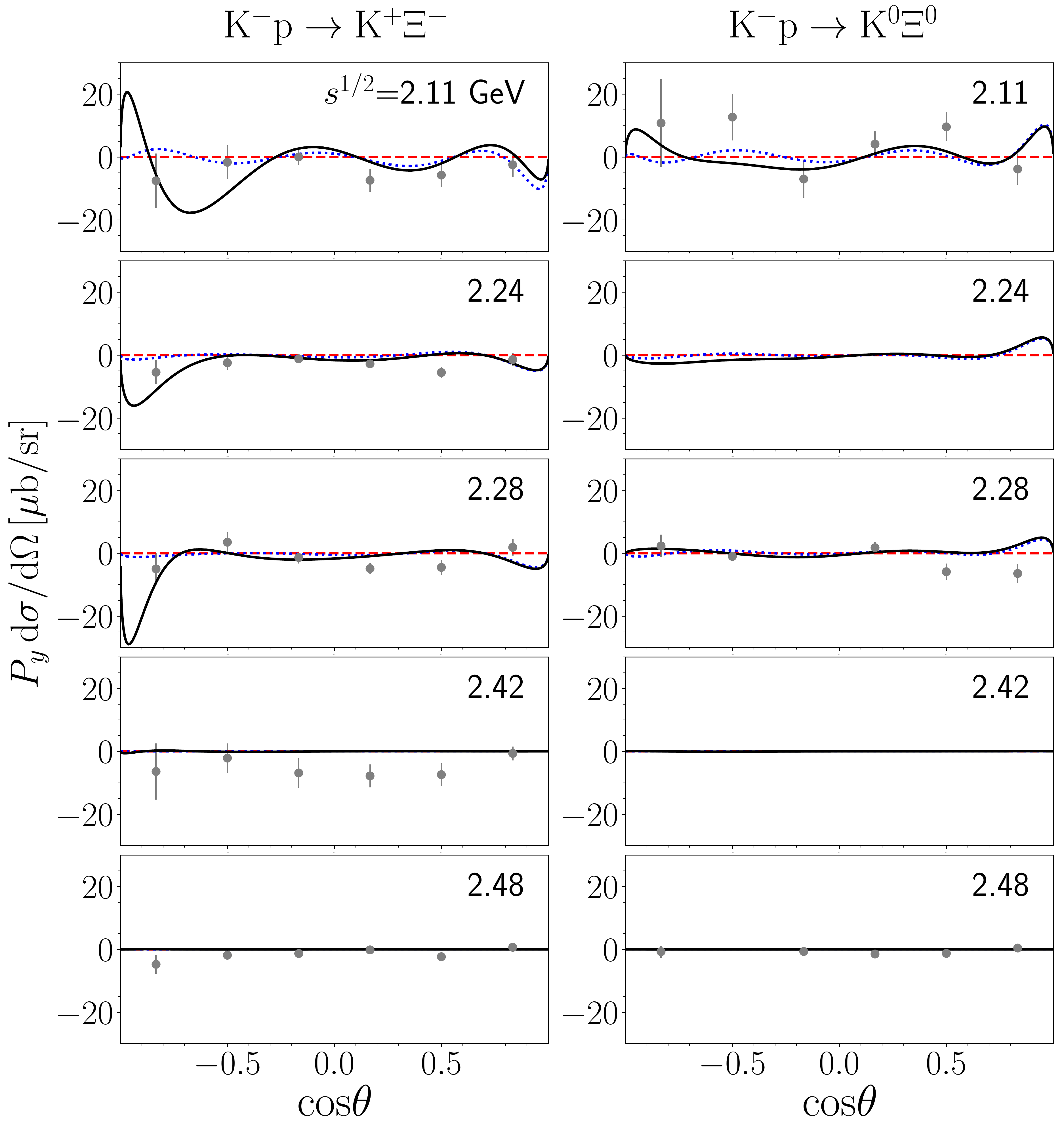}
\caption{Recoil asymmetries multiplied by differential cross sections,
$P_y \frac{d\sigma}{d\Omega}$, versus $\cos\theta$ for (a) $K^- p \to K^+ \Xi^-$ and
(b) $K^- p \to K^0 \Xi^0$ for different c.m. energies $\sqrt{s} = (2.11 - 2.48)$ GeV.
The curve notations are the same as those shown in Fig.~\ref{FIG05}.
The data are obtained from Refs.~\cite{Trippe:1967wat,Dauber:1969hg}.}
\label{FIG07}
\end{figure}
Fig.~\ref{FIG07} presents the results of the recoil asymmetries multiplied by the
differential cross sections, $P_y \frac{d\sigma}{d\Omega}$, as a function of $\cos\theta$
for (a) $K^- p \to K^+ \Xi^-$ and (b) $K^- p \to K^0 \Xi^0$.
In this study, the reaction plane is defined as $\hat{x}\times\hat{z}$ and the $z$-axis is
parallel to the direction of the $K^-$ beam.
The backward angles are significantly affected by the inclusion of the $s$-channel
$\Lambda$ and $\Sigma$ resonances for $K^- p \to K^+ \Xi^-$,
whereas changes in the forward angles were relatively mild in both channels.
The results of the recoil ($P_y$) and target ($T_y$) asymmetries are identical, that is,
$P_y = T_y$.
The other components vanish, that is, $P_i = T_i = 0$ for $i = x,z$ by
definition ~\cite{Jackson:2013bba}.

\subsection{Hybrid RPR model: Regge-plus-resonance model with
rescattering diagram}

\begin{figure}[htb]
\stackinset{c}{-1.cm}{t}{0.1cm}{(a)}{
\includegraphics[width=7.7cm]{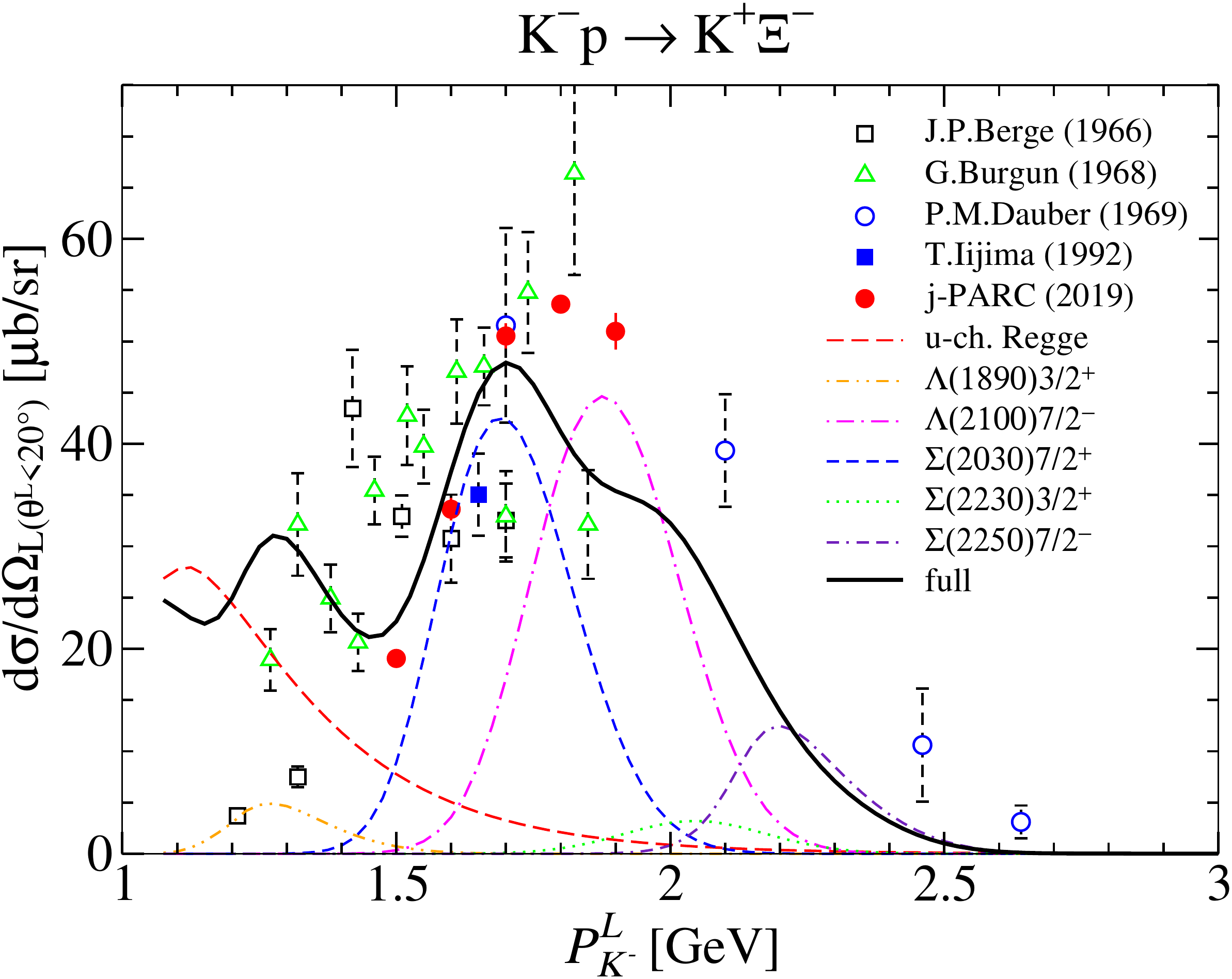}
}\\
\stackinset{c}{-1.cm}{t}{0.1cm}{(b)}{
\includegraphics[width=7.7cm]{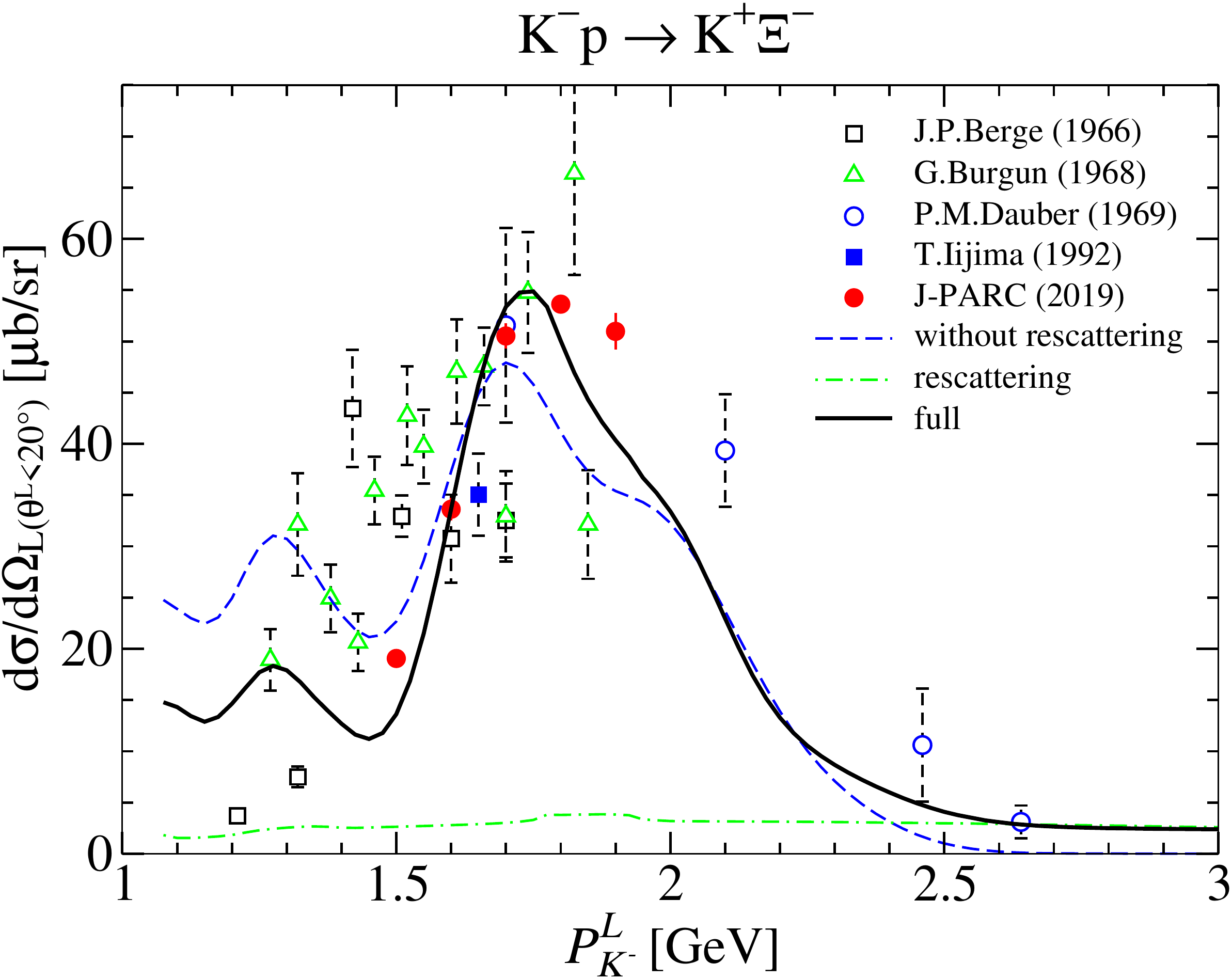}
}
\caption{Differential cross sections averaged over the forward region $\theta^L <
20^\circ$ as a function of the beam momentum for the $K^- p \to K^+ \Xi^-$ reaction 
(a) without and (b) with the rescattering diagram contribution.
(a) Curve notations are the same as those indicated in Fig.~\ref{FIG04}.
(b) The dashed blue line corresponds to the solid black line in the figure above.
The dot-dashed green line indicates the rescattering diagram contribution.
The solid black line indicates the full contribution including the rescattering diagrams.
The data are obtained from Refs.~\cite{Berge:1966zz,Burgun:1968ice,Dauber:1969hg,Iijima:1992pp,
Nagae:2019uzt}.}
\label{FIG08}
\end{figure}

We examined the laboratory cross sections averaged over the forward region 
for the $K^-p\to K^+\Xi^-$ reaction as a
function of the $K^-$ beam momentum, $p_{K^-}$, as shown in Fig.~\ref{FIG08}.
To combine the old bubble-chamber data with the recent datasets, we considered 
the angle average of the differential cross sections in the laboratory frame
($L$) \cite{Dover:1982ng}:
\begin{align}
\hskip-0.4cm\left\langle \frac{d\sigma}{d\Omega_{L}} \right\rangle_{\rm av} =
{\int_{\theta \rm min}^{\theta \rm max}} \frac{d\sigma}{d\Omega_{L}} d(\cos\theta^{L})
\left/
{\int_{\theta \rm min}^{\theta \rm max}} d(\cos\theta^{L}) \right..
\label{eq:DCSConvert}
\end{align}

The open squares, triangles, and circles shown in Fig.~\ref{FIG08} correspond to
the differential cross sections averaged over the laboratory angular region
$\theta^L < 18^\circ$.
These data points were deduced from the Legendre polynomial expansion of the differential
cross sections in Refs.~\cite{Berge:1966zz},~\cite{Burgun:1968ice}, and~\cite{Dauber:1969hg},
respectively.
The errors were scaled to the uncertainties of the total cross section. 
The closed squares and circles denote the cross section 
averaged over the laboratory angular regions $1.7^\circ < \theta^L < 13.6^\circ$~\cite{Iijima:1992pp} 
and $\theta^L <20^\circ$\cite{Nagae:2019uzt}, respectively.
 
We transformed the differential cross sections from the c.m. frame to the laboratory
frame by using the following kinematic relations: 
\begin{eqnarray}
\frac{d\sigma/d\Omega_{L}}{d\sigma/d\Omega} &=&
\frac{m_p |\mathbf{p}^{L}_{K^-}||\mathbf{p}^{L}_{K^+}|}{|\mathbf{p}_{K^-}| |\mathbf{p}_{K^+}|} 
\frac{1}{E_{K^-}^{L}+m_p- \dfrac{|\mathbf{p}^{L}_{K^-}|}{|\mathbf{p}^L_{K^+}|}E_{K^+}^{L}\cos\theta^{L}},\hskip+0.6cm\\
\cos\theta^{L} &=&
\frac{E_{K^-}^{L} E_{K^+}^{L} - E_{K^-} E_{K^+} 
+ |\mathbf{p}_{K^-}| |\mathbf{p}_{K^+}|
\cos\theta} {|\mathbf{p}^{L}_{K^-}| |\mathbf{p}^{L}_{K^+}|},\hskip+0.8cm 
\label{eq:FrameConvert}
\end{eqnarray}
where $(E^L,\mathbf{p}^L)$ and $(E,\mathbf{p})$ denote the energy and three-momentum vector
in the laboratory and c.m. frames, respectively.

The role of the rescattering diagrams is discussed in detail herein.
The individual contributions of the $u$- and $s$-channel diagrams are shown in
Fig.~\ref{FIG08}(a).
The two resonances, $\Lambda(2100)7/2^-$ and $\Sigma(2030)7/2^+$, are primarily responsible for
the peak near $p^L_{K^-}=1.8$ GeV$/c$, but are insufficient to reproduce the recent J-PARC
data~\cite{Nagae:2019uzt}.
The rescattering diagrams are additionally included
in Fig.~\ref{FIG08}(b).
A close inspection of our results using the J-PARC data verifies that the
rescattering diagrams are
an essential ingredient for the $K^- p \to K^+ \Xi^-$ reaction.
The total coherent sum of the amplitudes significantly changes with the inclusion of the
rescattering diagrams.
This is because the resonance contribution constructively interferes with
the rescattering diagrams at the laboratory scattering angles $\theta^L < 20^\circ$.
Certain descrepancies in the region $1.8 < p^L_{K^-} < 1.9$ GeV$/c$ can be filled by the
inclusion of high-mass meson-baryon channels, such as $V\Lambda^\ast(1405)$ and $V\Sigma^\ast(1385)$,
which is beyond the scope of this study.

We also observe that the $VB$ rescattering diagram [Fig.~\ref{FIG01}(c)] plays a major role,
whereas the $PB$ rescattering diagram [Fig.~\ref{FIG01}(d)] has a limited contribution to the
$K^- p \to K^+ \Xi^-$ reaction. Here, only the $T_{K^- p \to V B}^K T_{V B \to K^+ \Xi^-}^{K^\ast}$ 
and $T_{K^- p \to V B}^{K^\ast} T_{V B \to K^+ \Xi^-}^K $ processes 
were considered among the $VB$ rescattering diagrams.
In contrast, the $T_{K^- p \to V B}^K T_{V B \to K^+ \Xi^-}^K$ and $T_{K^- p \to V B}^{K^\ast}
T_{V B \to K^+ \Xi^-}^{K^\ast}$ processes are highly suppressed.

\begin{figure}[ht]
\includegraphics[width=8.0cm]{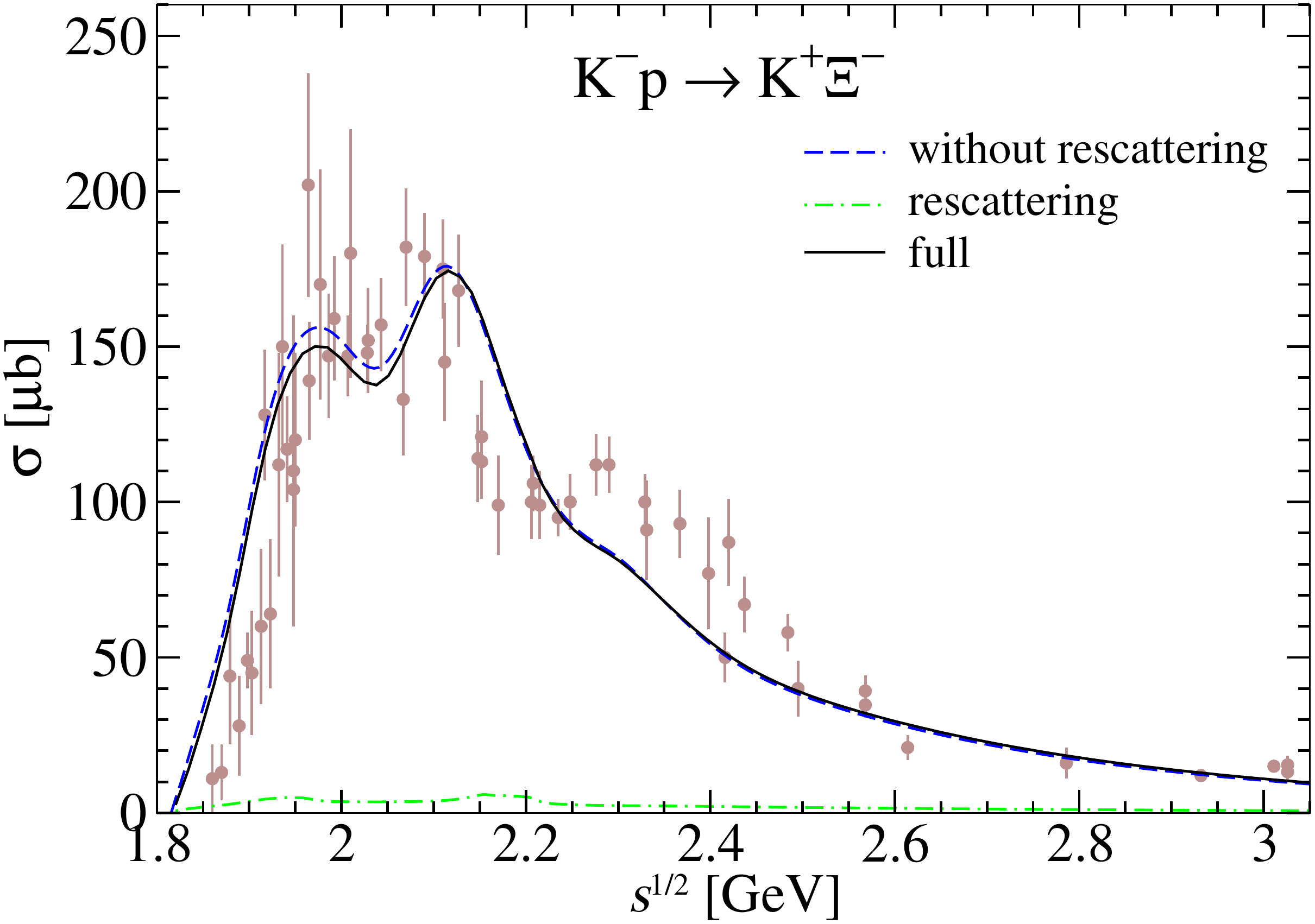} \hspace{1em}
\caption{Total cross sections as a function of the c.m. energy $\sqrt{s}$ 
for the $K^- p \to K^+ \Xi^-$ reaction.
Curve notations are the same as those in Fig.~\ref{FIG08}(b).
The dashed blue line indicates the contribution of the $u$- and $s$-channel diagrams, and
the dotted-dashed green line indicates the rescattering diagram contribution.
The solid black line indicates the full contribution including the rescattering diagrams.}
\label{FIG09}
\end{figure}
Fig.~\ref{FIG09} depicts the calculated total cross sections as a function of $\sqrt{s}$ for
the $K^- p \to K^+ \Xi^-$ reaction when the rescattering diagrams are added.
The contribution of the $u$- and $s$-channel diagrams
destructively interferes with the rescattering diagram contribution in the region $\sqrt{s} < 2.15$ GeV.

\begin{figure}[ht]
\stackinset{l}{1.cm}{t}{0.2cm}{(a)}{
\includegraphics[width=4.30cm]{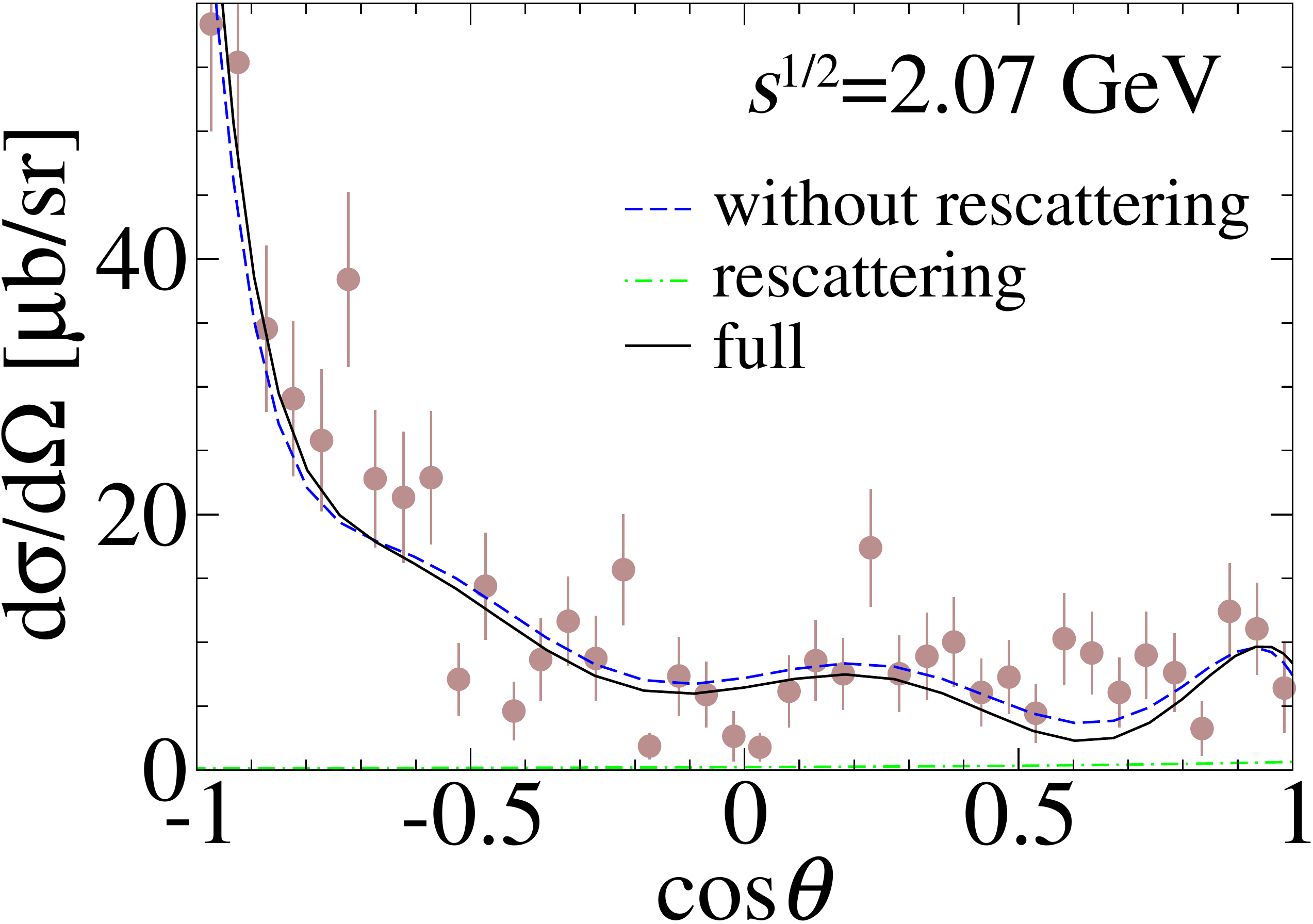}
}\stackinset{l}{0.6cm}{t}{0.2cm}{(b)}{
\includegraphics[width=4.05cm]{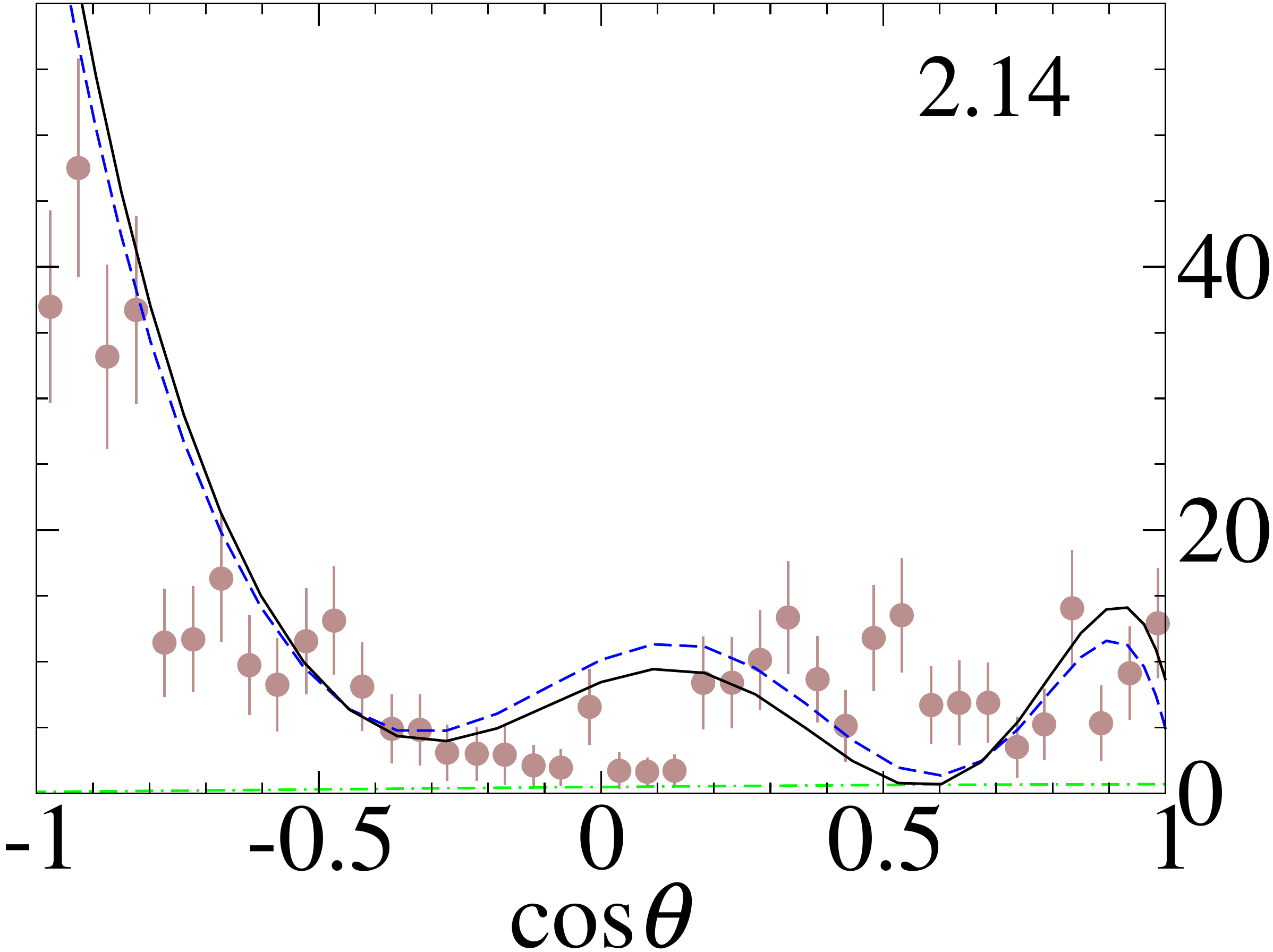}
} \hspace{1.3em}
\stackinset{l}{1.cm}{t}{0.2cm}{(c)}{
\includegraphics[width=4.30cm]{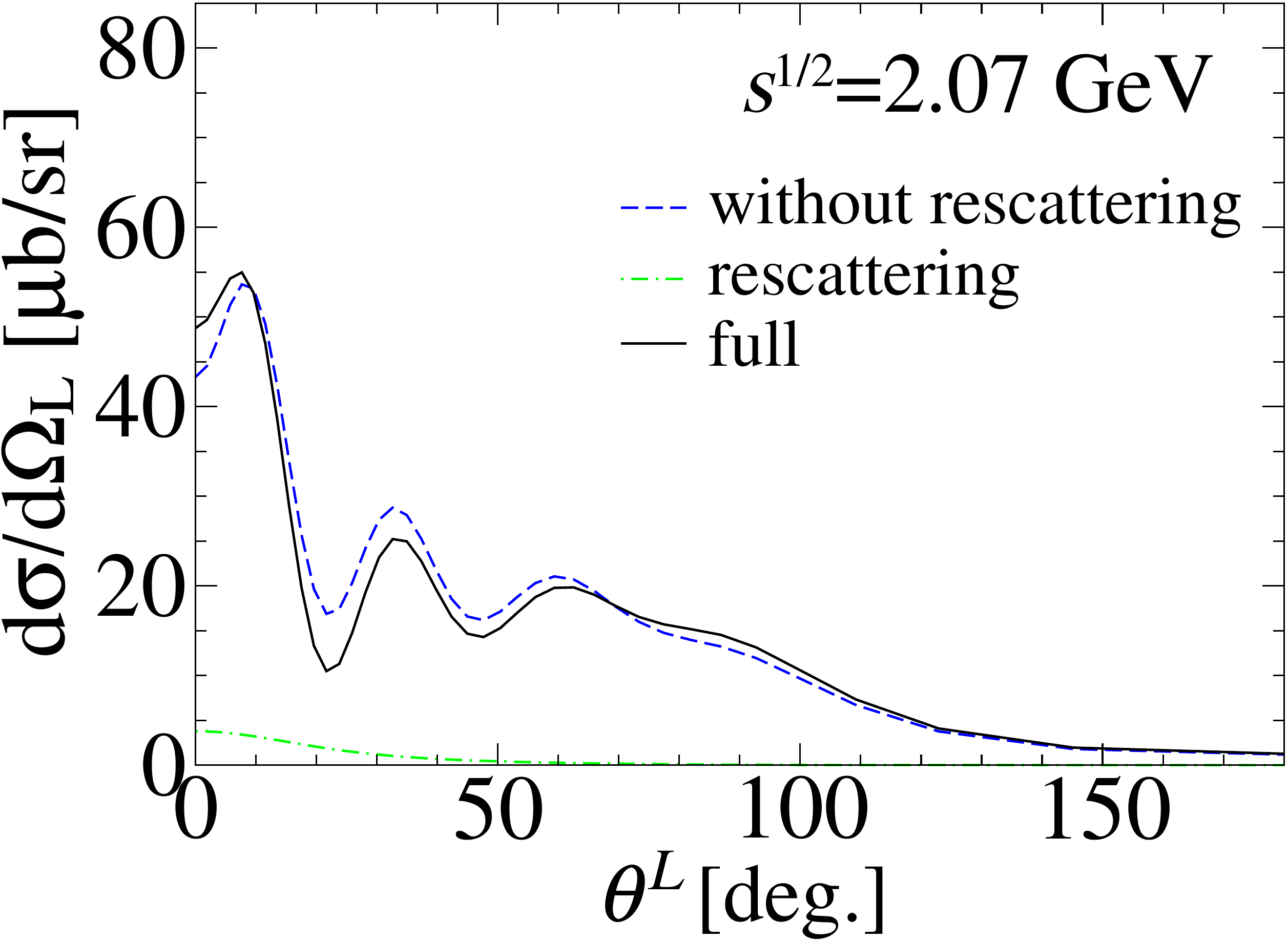}
}\stackinset{l}{0.6cm}{t}{0.2cm}{(d)}{
\includegraphics[width=4.05cm]{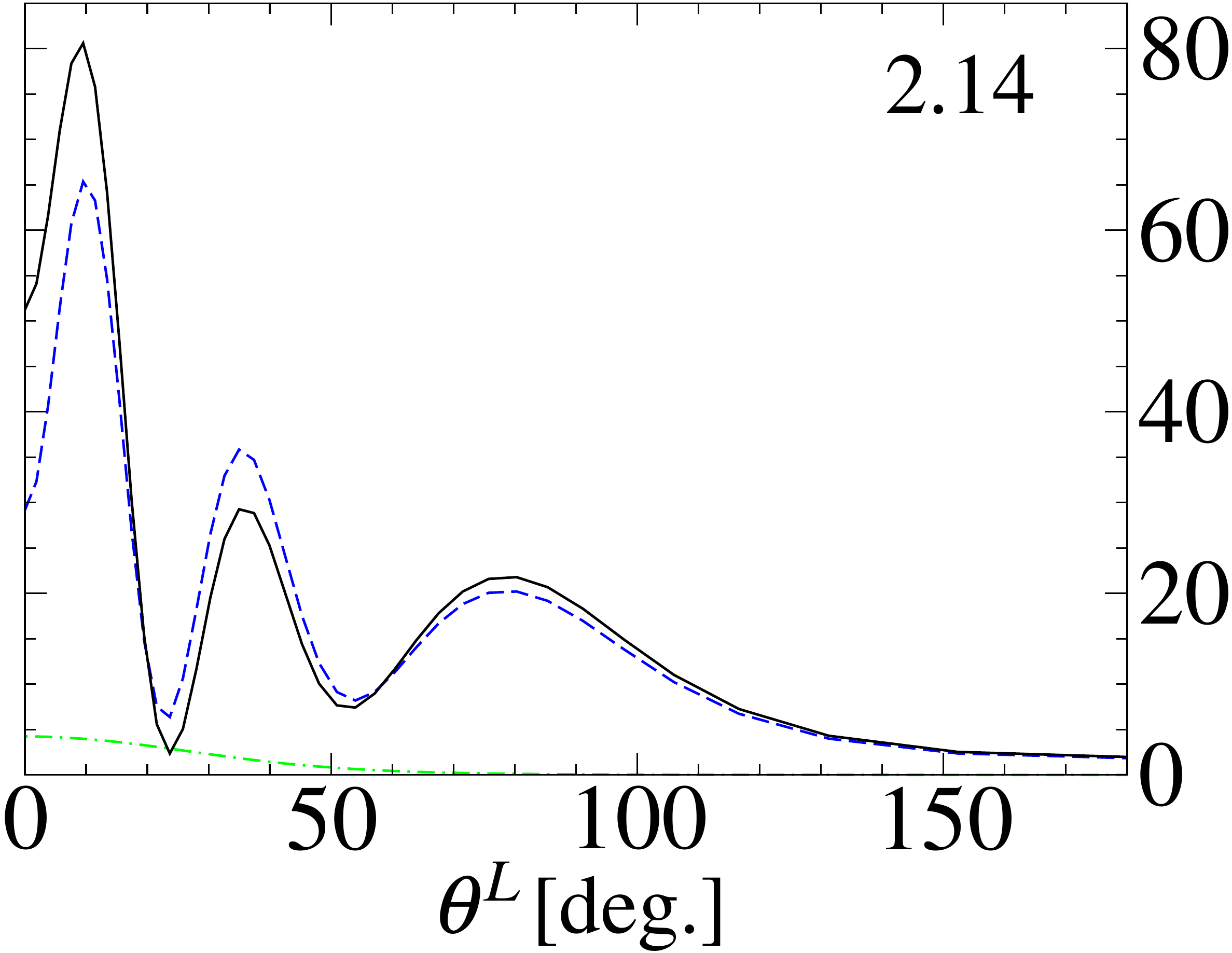}
}
\caption{Differential cross sections for $K^- p \to K^+ \Xi^-$ in the c.m. (a,b) and
laboratory (c,d) frames at $\sqrt{s} = 2.07$ (a,c) and $\sqrt{s} = 2.14 $ (b,d) GeV.
Curve notations are the same as those in Fig.~\ref{FIG08}(b).}
\label{FIG10}
\end{figure}
In Fig.~\ref{FIG10}, we present the calculated differential cross sections for the $K^- p
\to K^+ \Xi^-$ reaction in the c.m. (a,b) and laboratory (c,d) frames at $\sqrt{s} = 2.07$ (a,c) and
$\sqrt{s} = 2.14$ (b,d) GeV when the rescattering diagram is added.
A sharp decrease in the forward angle region was clearly observed at $\sqrt{s} = 2.14 $ GeV.
The more $\Lambda(2100)7/2^-$ and $\Sigma(2030)7/2^+$ overlap, the steeper the contribution
of the $s$-channel diagram decreases at forward angles.

Finally, the rescattering diagrams have a limited effect on the
$K^- p \to K^0 \Xi^0$ reaction.
They constructively interfere with the contributions of the $u$- and $s$-channel diagrams
and their inclusion worsens the results.

\section{Summary} \label{Section:IV}

We investigated the double-charge and double-strangeness exchange reactions $\overline{K}N \to K\Xi$
in the hybrid Regge-plus-resonance model involving the rescattering diagrams.
For the background contributions, we considered the $\Lambda$, $\Sigma$, and
$\Sigma(1385)$ Regge trajectories in the $u$ channel; $\Lambda$ and $\Sigma$ ground states in the $s$ channel, 
which are responsible for explaining the backward peaks.
In addition, various $\Lambda$ and $\Sigma$ hyperon resonances were considered
in the $s$-channel diagram to account for the bump structures
of the total cross sections in the low energy region,
$\sqrt{s}< 2.5$ GeV.
Based on the known branching ratios for the $Y^\ast \to\overline{K} N$ decays, we extracted the
branching ratios for the $Y^\ast \to K \Xi$ decays by fitting with the data, where
$Y^\ast$ denotes $\Lambda$ or $\Sigma$ resonances.

We found that the $\Sigma$ and $\Sigma(1385)$ Regge trajectories are crucial in the $K^- p \to
K^0 \Xi^0$ reaction, and the $\Lambda$ Regge trajectory is predominant in the $K^- p \to K^+
\Xi^-$ reaction.
Furthermore, the $\Lambda(2100)7/2^-$ and $\Sigma(2030)7/2^+$ resonances were
the most significant in the $s$ channel.
The $\Lambda(1890)3/2^+$, $\Sigma(2230)3/2^+$, and $\Sigma(2250)$ resonances exhibited
sizable effects, assuming that the latter are $J^P = 7/2^-$.

The roles of $\Lambda(2100)7/2^-$ and $\Sigma(2030)7/2^+$ in the $s$ channel are
clarified more accurately when we examine the differential cross sections.
The results for the resonance region $2.07<\sqrt{s}< 2.15$ GeV at forward angles
were entirely different when both channels were compared.
The differential cross sections drastically decrease at significantly forward angles ($\cos\theta
> 0.9$) in the $K^+\Xi^-$ channel, but sharply increase in the $K^0\Xi^0$ channel.
The different isospin combinations of the $\Lambda(2100)7/2^-$ and $\Sigma(2030)7/2^+$
amplitudes cause the two distinct behaviors.

We present the results of the recoil asymmetries multiplied by the differential cross sections, 
$P_y d\sigma/d\Omega$, for both channels.
The results for $K^- p \to K^+ \Xi^-$ at backward angles are sensitive to the inclusion of the
$s$-channel $\Lambda$ and $\Sigma$ resonances.
Our model predictions were also applied to the $K^- n \to K^0 \Xi^-$ reaction, and the total
and differential cross sections sufficienctly agreed 
with the available experimental data.

We considered the meson-baryon rescattering
diagrams for the $K^- p \to K^+ \Xi^-$ reaction. 
The $(\phi,\rho,\omega)$-$(\Lambda,\Sigma)$ rescattering diagram presents dominant
contribution, whereas the contribution of the $(\pi,\eta)$-$(\Lambda,\Sigma)$ rescattering
diagram is nearly negligible.
The rescattering diagram contribution destructively interfere with the
contribution of the $u$- and $s$-channel diagrams in the region $\sqrt{s} < 2.15$ GeV.
Our results were improved by including the rescattering diagrams for reproducing
the recent J-PARC data of the forward differential cross sections.
Future experiments at the J-PARC facility can verify our model predictions.

This study is the first step toward developing reasonable reaction theories of meson-induced reactions.
The extension of our hybrid model to other $\pi$- or $K$-induced reactions is
essential for understanding the relevant reaction mechanisms more systematically, 
and will significantly contribute to the development of baryon spectroscopy;
relevant research is currently in progress.


\acknowledgments
The study was supported by the Basic Science Research Program of the National Research
Foundation of Korea (NRF) under Grant Nos.~2021R1A6A1A03043957 (S.-H.K. and M.-K.Ch.), 
2020R1A2C3006177 (S.-H.K. and M.-K.Ch.), 2018R1A5A1025563 (S.i.N. and J.K.A.), 
2022R1A2C1003964 (S.i.N), and 2022K2A9A1A0609176 (S.i.N).



\end{document}